\documentclass[aps,prd,reprint,superscriptaddress,amsmath, amssymb]{revtex4-1}

\usepackage{graphicx}
\usepackage{siunitx}
\usepackage[mathlines]{lineno}
\usepackage{amssymb}
\usepackage{amsmath}
\usepackage{comment}
\usepackage{soul, xcolor}

\begin{document}

\preprint{APS/123-QED}

\setstcolor{red}
\newcommand\ly[1]{{{#1} }}

\title{Neutron Reconstruction via Blips in Liquid Argon Time Projection Chambers}

\author{Miguel Hernandez Morquecho}
\email{hernanm@umn.edu}
\affiliation{%
  University of Minnesota, USA
}%
\author{Bryce Littlejohn}
\email{blittlej@illinoistech.edu}
\affiliation{%
  Illinois Institute of Technology, USA
}%

\author{Paola Sala}
\email{psala@fnal.gov}
\affiliation{%
  Particle Physics Department, Fermilab, P.O. Box 500, Batavia, Illinois 60510, USA
}%

\author{Linyan Wan}
\email{lwan@fnal.gov}
\affiliation{%
  Particle Physics Department, Fermilab, P.O. Box 500, Batavia, Illinois 60510, USA
}%

\date{\today}

\begin{abstract}

	Neutrons are important final-state particles in neutrino interactions, yet they are not considered or reconstructed in most current neutrino LArTPC physics analyses. 
	In this paper, we present a simulation-based proof-of-concept study of neutron reconstruction in a generic LArTPC detector.  
	Leveraging isolated, MeV-scale energy deposits, or blips, from neutron inelastic scattering, and using realistic blip response from published experimental results, we demonstrate the capability to identify neutrons and to reconstruct the direction and energy of the final-state neutron system in sub-GeV neutrino interactions.  
	We then explore how neutron-related blip attributes can be used to improve physics studies of neutrino interactions, such as enhancing neutrino-antineutrino separation in atmospheric neutrinos and reverse-horn-current beam neutrinos.  
    This study provides an initial quantification of LArTPC neutron reconstruction capabilities, which we expect to improve with future advancements in blip reconstruction, identification, and classification algorithms, as well as the modeling of neutrons.
\end{abstract}

\maketitle
%\linenumbers

\section{Introduction}

As neutrino oscillation measurements enter the precision era~\cite{DayaBay:2022orm, JUNO:2025gmd, T2K:2025wet, DUNE:2020jqi, Hyper-Kamiokande:2025fci}, it becomes crucial that we understand neutrino interaction modeling and constrain the significant uncertainties arising from nuclear effects. 
Modeling these nuclear effects is particularly critical for sub-GeV neutrino interactions, where the energy transfer in the neutrino interaction can be comparable to or dominated by Fermi motion, nucleon-nucleon correlations, and final-state interactions.
These sub-GeV neutrinos bear important physics potential, to name a few: CP violation via sub-GeV atmospheric neutrinos~\cite{Kelly:2019itm, Kelly:2023ugn} and beam neutrinos~\cite{Bass:2013vcg,Blennow:2019bvl,Hyper-Kamiokande:2025fci}, sterile neutrino searches via beam neutrinos~\cite{Machado:2019oxb}, and cross-section studies towards constraining uncertainties in neutrino oscillation studies~\cite{SBND:2025lha, MicroBooNE:2019nio, Coyle:2025xjk}.
Outgoing hadrons serve as valuable probes of these effects, among which protons are much more well-constrained from experimental data~\cite{ArgoNeuT:2014ihi,MicroBooNE:2018xad,T2K:2018rnz, MicroBooNE:2020akw, MicroBooNE:2022emb, MicroBooNE:2024yzp, MicroBooNE:2024zkh, MicroBooNE:2024zwf,NOvA:2024zmr} compared to uncharged neutrons, due to the difficulties involved in their detection.
In neutrino interactions, a significant amount of momentum in 1 GeV neutrinos is carried by neutrons. 
Neutron reconstruction therefore possesses crucial importance to recover substantial information loss in the initial neutrino kinematics~\cite{Friedland:2018vry,Castiglioni:2020tsu}.

In neutrino detectors using water Cherenkov or liquid scintillation technology, neutrons typically can be identified and reconstructed via capture on a target nucleus such as hydrogen~\cite{LSND:1996jxj, Suzuki:2014woa,DoubleChooz:2019qbj,DayaBay:2024hrv, Super-Kamiokande:2025cht} or other doped elements~\cite{cowan1956, Abbes:1995nc, CHOOZ:2002qts, DayaBay:2012fng, Bellerive:2016byv, PROSPECT:2018dnc, Super-Kamiokande:2022cvw}.
In a liquid argon time projection chamber (LArTPC), however, due to the destructive interference feature in neutron-argon cross-section at $\sim$\SI{57}{keV}~\cite{Brown:2018jhj, ARTIE:2022wqs}, low-energy neutrons in liquid argon have an interaction length of $\sim$\SI{30}{m}, indicating that neutron capture on argon, although yielding high energy $\gamma$-rays with detectable signatures, is less likely to be contained within a LArTPC detector.
Instead, the detection of neutrons can be done via their inelastic scattering off of argon nuclei.

Neutrons from neutrino interactions can inelastically scatter with argon nuclei, leaving the nuclei in excited states and potentially producing secondary neutrons and protons~\cite{MacMullin:2012fz,Andringa:2023aax}. 
The de-excitation of the remnant nucleus produces $\gamma$-rays of a few MeV, which, through Compton scattering and electron-positron pair production, show up in LArTPC data as isolated charge depositions called blips. 
Several LArTPC experiments have reported evidence of neutron identification, including ArgoNeuT~\cite{ArgoNeuT:2018tvi} using blips and MicroBooNE~\cite{MicroBooNE:2024hun} using displaced secondary protons. 

In this paper, we conduct a phenomenological study of how neutronic signatures can be implemented to improve high-level neutrino LArTPC physics deliverables.  
We focus especially on deliverables relevant to sub-GeV neutrinos, where the neutron-carried information is more dominant, yet less studied.
Neutron-induced proton emission is much less frequent in this energy range: for example, MicroBooNE's recent study focusing on secondary proton identification reports an efficiency plateauing at $\sim$3\% around \SI{250}{MeV} neutron kinetic energy~\cite{MicroBooNE:2024hun}.  
This study will therefore primarily center on neutron reconstruction through the use of $\gamma$-induced blips.  

Our scheme reconstructs the presence, direction, and energy of neutronic final states using a simulated sample with realistic experimental response of blips, which are then purified via a set of simple analysis cuts.  
These simple neutron attributes are found to improve charge-sign purities for sub-GeV atmospheric and beam neutrino interaction samples.  
The outcome represents useful starting benchmarks for understanding the utility of blip-based neutron reconstruction in neutrino LArTPCs which can be expected to improve with more advanced experimental techniques in signal blip reconstruction and selection schemes.  

This paper is organized as follows.  
We first describe the study setup in Section~\ref{sec:sim} including neutrino interaction modeling, particle transport simulation, and detector response and reconstruction for blips.  
Section~\ref{sec:bkgred} talks about the categorization of blips and the analysis selections to reject non-neutron blip background.
Neutron reconstruction performance is highlighted in Section~\ref{sec:neutronrec}, including neutron identification, energy reconstruction, and direction reconstruction.  
Section~\ref{sec:physapp} gives a few examples of physics applications using this neutron reconstruction technique, which demonstrates potential enhancement in neutrino-antineutrino separation tasks in multiple neutrino sources.  
Section~\ref{sec:discuss} discusses systematic uncertainties and the future outlook.
  Takeaways are summarized in Section~\ref{sec:summary}.  

\section{Simulation}
\label{sec:sim}

The study is set up using sub-GeV atmospheric neutrino flux predicted by the Honda model~\cite{Honda:2015fha}.  
For our baseline neutrino samples, final-state particle generation and particle transport is conducted using \texttt{FLUKA}~\cite{Ballarini:2024isa}, as LArTPC experiments have so far observed best agreement between neutron-related datasets and \texttt{FLUKA} simulation~\cite{ArgoNeuT:2018tvi}. 
To explore related systematic uncertainties, in Section~\ref{sec:discuss} we also investigate alternative neutrino interaction modeling using \texttt{GENIE}-hA~\cite{Andreopoulos:2009rq, GENIE:2021zuu} and \texttt{GENIE}-INCL~\cite{Liu:2026wlw}, and alternative particle transportation using the \texttt{Geant4} package~\cite{GEANT4:2002zbu}.  
While neutrons also play an important role in neutral current interactions, in this analysis we focus on charged current (CC) interactions to provide information and inputs to oscillation analyses. 

\subsection{Neutrino Event Generation}

\texttt{FLUKA} has its own neutrino interaction generator that embeds the basic neutrino-nucleon interaction in a \texttt{FLUKA} nuclear model called \texttt{PEANUT}~\cite{Ballarini:2024isa}.
Quasi-elastic neutrino interactions are sampled according to the Llewellyn Smith formalism~\cite{Llewellyn}, accounting for the mass of the produced lepton.  
At higher energies, the resonant and deep inelastic scattering (DIS) channels open. 
The production of $\Delta$ resonances is modeled following Rein-Sehgal~\cite{Rein-Sehgal}.
No other resonance is considered. 
Care is taken to avoid double-counting with the non-resonant DIS channel, with a linear decrease (increase) of the resonance (DIS) probability as a function of the mass of the intermediate hadronic system.
The (anti)neutrino-nucleon DIS generator, called NUNDIS, is described in ref.~\cite{NUNDIS2}.  
However, due to the neutrino energy range considered in this work, the DIS contribution is close to negligible.  

Following these neutrino interaction cross-section starting points, the \texttt{PEANUT} model within \texttt{FLUKA} provides initial and final state interactions. 
The initial state is described as a Fermi gas with Fermi momentum dependent on the local nuclear density. 
A Pauli blocking check is performed at each interaction. 
The coherence length and formation zone are applied to the interaction products to account for the indeterminacy of the location of the products.  

The interaction products are then transported in the nucleus according to a generalized intranuclear cascade, followed by a pre-equilibrium stage and then by equilibrium emission of particles and fragments.  
Any residual excitation energy is dissipated via the emission of $\gamma$-ray cascades.  
Energy conservation is ensured at each step, exploiting experimental and evaluated data on nuclear masses.
$\gamma$ de-excitation is based as far as possible on existing experimental data on nuclear levels and branching ratios~\cite{RIPL}.  
Outside the coverage of databases, a native statistical model is applied~\cite{ZPhysC71}.  
A validation of the algorithm, with specific focus on the Fermi motion as implemented in PEANUT, is presented in an experimental paper on neutrino interactions in liquid argon~\cite{50l}.  
The good agreement of \texttt{FLUKA} simulations with nuclear de-excitation $\gamma$ data collected in liquid argon is presented in Ref.~\cite{ArgoNeuT:2018tvi}. 

\begin{figure}[!ht]
   \centering
         \includegraphics[width=0.89\linewidth]{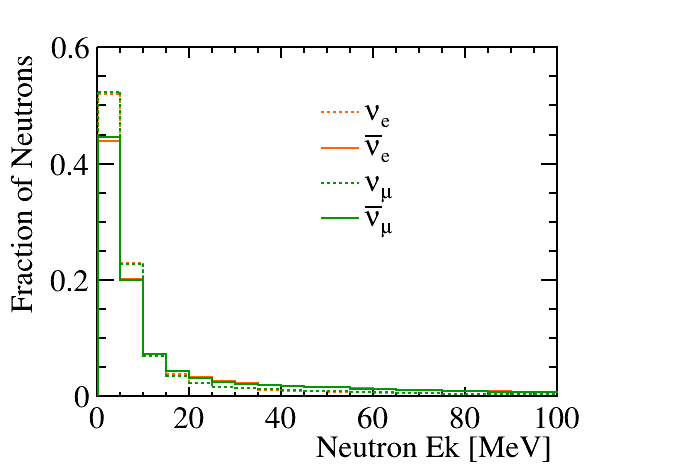}
         \includegraphics[width=0.89\linewidth]{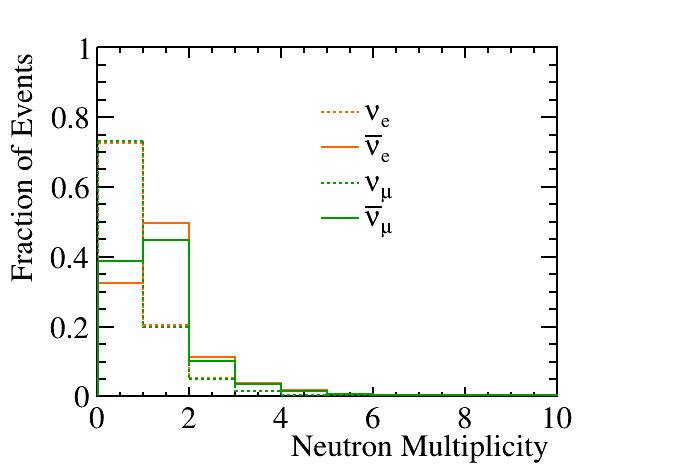}
			\caption{Neutron kinetic energy distribution with a \SI{2}{MeV} threshold (top) and neutron multiplicity distribution with a \SI{5}{MeV} threshold (bottom) from sub-GeV atmospheric neutrinos generated using \texttt{FLUKA}.}
			\label{fig:fluka_neutron}
\end{figure}

Figure~\ref{fig:fluka_neutron} shows the energy distribution for primary neutrons and the primary neutron multiplicity from the \texttt{FLUKA} event generation for sub-GeV atmospheric neutrino charged current interactions.
The energy distribution has a \SI{2}{MeV} threshold to avoid the overwhelmingly large evaporation neutron population.
This choice is also comparable with prior neutron-related measurements from beam neutrino experiments.~\cite{MINERvA:2019wqe}. 
The neutron multiplicity figure has a \SI{5}{MeV} threshold, as required by neutron visibility in a LArTPC, discussed in Section~\ref{subsec:neutronid} and Section~\ref{subsec:neutprod}.  

\subsection{Particle Transport}
\label{subsec:partran}

Following the generation of final-state particles of neutrino-argon interactions, particles must be transported through the liquid argon medium.  
Final-state products are generated at a single central point in a large homogeneous 4.0$\times$4.0$\times$5.0~m$^3$ LAr volume, with secondary interaction products above a minimum tracking threshold generated and propagated throughout the entire volume.  
This LAr volume is comparable in scale to that of the SBND, ICARUS, and ProtoDUNE LArTPC detectors~\cite{SBND:2025lha,ICARUS:2023gpo,protodunejinst} and to the size of one TPC within a DUNE far detector 10~kT module~\cite{dune2020}.  
When products reach the edge of this simulated volume, they cease being tracked or propagated.  

We perform the particle transport in the baseline sample in two stages: \texttt{FLUKA} particle transport, and \texttt{LArSoft} electron transport.
For \texttt{FLUKA} particle transport, we select transport settings that prioritize accuracy over reduced CPU usage, such as enabling the production of heavy fragments during evaporation and enabling photo-nuclear interactions.  
In particular, the transport of neutrons below 20~MeV is performed with the recent fully correlated point-wise algorithm~\cite{Satif15_neutrons} down to thermal energies. 
This ensures that $\gamma$ cascades following neutron interactions are faithfully reproduced, and that correlations among the interaction products are conserved.  

Most neutrino LArTPC simulation and analysis studies perform final-state particle transport using \texttt{Geant4} implemented within the larger structure of the \texttt{LArSoft} LArTPC analysis framework~\cite{Snider:2017wjd} (see Refs.~\cite{MicroBooNE:2017xvs,MicroBooNE:2018xad,Castiglioni:2020tsu} for useful descriptions).  
Since \texttt{FLUKA} is not directly interfaced with \texttt{LArSoft}, we implemented a scheme to transfer \texttt{FLUKA}-generated particles into \texttt{LArSoft}.   
This transfer is performed at the electron transport phase, with electrons produced but not transported within \texttt{FLUKA}.
Truncated electrons are instead propagated within \texttt{LArSoft} and \texttt{Geant4}.  
Their initial position, momentum, and energy are used as input conditions for transport within the LAr volume in \texttt{LArSoft}, with information from all primaries and secondaries retained for downstream processing and analysis.  
This approach reduces computational cost while allowing for generation of reconstructed blips in \texttt{LArSoft} with similar data structures for both particle transport paths.  

As a cross-check of the two-stage implementation and as a particle transport systematic exercise, we also perform particle transport in an alternative \texttt{Geant4} setup.  
We use the Bertini cascade model~\cite{Heikkinen:2003sc} to simulate hadron–nucleus interactions at intermediate energies, producing secondary particles such as neutrons. 
Neutrons with energies below 20 MeV are treated using the \texttt{NeutronHP} library~\cite{Plompen:2020due}, a data-driven approach based on evaluated nuclear datasets. 
These cross-check results are presented in detail in the Supplementary Material~\cite{appendix} and are also described in Section~\ref{subsec:neutprop}.  

\subsection{Blip Reconstruction}
\label{sec:bliprec}

Energy depositions from \texttt{FLUKA}-based and \texttt{Geant4}-based particle transport paths are processed by \texttt{BlipReco}, a \texttt{LArSoft}-native reconstruction package first described in Ref.~\cite{Castiglioni:2020tsu} and implemented in recent MicroBooNE MeV-scale analyses~\cite{MicroBooNE:2023sxs,MicroBooNE:2024prh}.  
The \texttt{BlipReco} package is primarily designed to reconstruct physics quantities of interest from either simulated or real LArTPC wire signals generated by isolated MeV-scale energy depositions, or blips, in LArTPCs.  
In addition, it processes simulated particle history outputs from \texttt{LArSoft} to generate truth-level physics quantities mirroring those reconstructed from wire signals.  
Lastly, \texttt{BlipReco} generates truth-level quantities related to the origins of reconstructed blips, such as the identity of the particle that generated a blip's ionization.  
In this analysis, we avoid simulation and processing of TPC wire signals and leverage only truth-level quantities from \texttt{BlipReco}, similarly to Ref.~\cite{Castiglioni:2020tsu}.  

When performing blip signal selection and analysis we make use of truth-level analogs to two primary reconstructed blip quantities: energy and 3D position.  
The true deposited energy associated with a blip is formed by summing all true energy deposited as ionization between the start and end of a charged particle's tracked life.  
Ionization generated by secondary but connected charged particle trajectories must be added to this sum, while energy lost via radiative ($\gamma$-emitting) processes must be subtracted from it.  
True 3D positions are defined as the midpoint between contiguous ionization start- and end-points. 
Truth-level blip formation is restricted to low-energy electrons ($0-12$ MeV kinetic energy) and protons ($0-40$ MeV kinetic energy).  

To more realistically approximate the response of a real LArTPC to MeV-scale charge depositions for this study, we apply blip response matrices provided by MicroBooNE in Ref.~\cite{MicroBooNE:2024prh}.  
Each reconstructed blip has its true energy smeared by randomly sampling from the probability density function contained within the response matrix for that true energy value.  
For blips generated by light or intermediate charged particles ($e^{\pm}, \pi^{\pm}, \mu^{\pm}$), MicroBooNE's electron response matrix is applied, while for heavy charged particles ($p$), its proton response matrix is used.  
Energy-dependent efficiency losses are accounted for similarly by randomly sampling MicroBooNE's efficiency response for the relevant true energy value.  
Since MicroBooNE blip efficiencies only approach a maximum of 80-90\% due to the presence of dead wires in their TPC, we optimistically re-scale this efficiency to 100\% in the $>$MeV regime.  
For blips truly generated by electrons, we optimistically apply the efficiency matrix corresponding to the MicroBooNE's collection plane, which achieves a reconstruction efficiency of $>$50\% of its maximum value around 200-250~keV of electron-equivalent energy deposition.  

The truth-level blip ancestry variables used in this analysis are the identity of the particle that generated a blip's ionization, the identity of the primary (non-neutrino) upstream ancestor of the blip's producing particle, and the process names of the particle interactions that generated upstream ancestors of interest. 
Since a large fraction of the \texttt{FLUKA} particle transport process happens outside of \texttt{LArSoft}, care was taken to properly record and parse particle histories in a similar manner for both simulation pathways.  
When multiple particles with differing identities (such as an $\gamma$-induced $e^{+}e^{-}$ pair) contribute to a blip's energy deposition, the particle contributing the most energy is defined to be the true originator.  
Determination of a blip's assigned primary upstream ancestor involves backtracking step-by-step through a blip's particle parentage until one reaches a primary particle produced directly by the neutrino event generator.  
There are a few exceptions to this ancestor-assignment process:
 
\begin{itemize}
\item{\underline{Stopped final-state $\mu$ products}: If a blip is made by the product of a final-state $\mu^-$ nuclear capture, the blip can sometimes be identified not as a `lepton-ancestor' blip, but is instead is associated with the upstream parent generated by the $\mu^-$ terminating process.  Specifically, blips generated by a $\mu^-$ nuclear capture neutron or a de-excitation $\gamma$-ray are termed `$n$-ancestor' blips and `$\gamma$-ancestor' blips, respectively.
}

\item{\underline{Secondary $n$ products}: If a blip is made by the products of a secondary neutron, it will be identified as a `$n$-ancestor' blip.  However, the `non-primary' nature of this neutron ancestor will also be recorded in some cases.  Blips generated by secondary neutrons with an ultimate final-state neutron ancestor retain a `primary' $n$-ancestor designation.}

\item{\underline{$\pi^0$ decay products}: In this study, we will disregard all neutrino interactions including a true primary or secondary $\pi^0$.  Their inclusion would require additional selection criteria, as well as the modeling of electromagnetic shower topologies in the $\nu_\mu$ and $\bar{\nu}_\mu$ samples.}  
\end{itemize}

Since wire simulation and other high-level reconstruction stages are not implemented, reconstructed track and shower information is not available for this study.  
Instead, we rely on truth-level energy deposition and ancestry information provided in \texttt{BlipReco} particle histories to relate blips to higher-energy topological features in a neutrino interaction.  
For example, prior to the blip selection and categorization process, we strike all $\delta$-rays from consideration by rejecting low-energy electrons with a direct $\mu^{\pm}$ parent and an ionization-related creation process.  

\section{Signal and Background Blip Categories}
\label{sec:bkgred}

To reconstruct neutron information in a neutrino interaction with blips, the first step is to identify blips with primary neutron origins from other neutrino-related and ambient blip backgrounds.
This section describes the features of blips from different origins, including the primary neutron signal, and backgrounds including non-primary neutrons, other blips from neutrino interactions, and blips from non-neutrino sources.

\subsection{Backgrounds from Final-State Leptons}
\label{subsec:lepton}

A significant fraction of blips in charged-current neutrino interactions have lepton-associated origins.  
For electron-flavored charged-current neutrino interactions, the primary lepton-associated blips comprise $77\%$ ($75\%$) of the total count from $\nu_e$ ($\bar\nu_e$) charged current interactions via electromagnetic showers, and it is usually not feasible to cluster all the energy deposition from the primary electron or $\gamma$-rays into the reconstructed shower objects.  
Muons at sub-GeV energy do not create these showers, but they can release $\delta$-ray $e^-$ from argon atoms via hard scattering, which deposit track-adjacent activity sometimes reconstructed as blips in real LArTPC data.  
In addition, $\delta$-rays can experience radiative energy losses, with bremsstrahlung photons generating blip activity disconnected from, but nearby, a $\mu$ track.
In $\nu_\mu$ ($\bar\nu_\mu$) charged current interactions, blips from this radiative process and from Michel electrons account for $14\%$ ($37\%$) of the total count.  

\begin{figure}[!ht]
    \centering
    \includegraphics[width=0.89\linewidth]{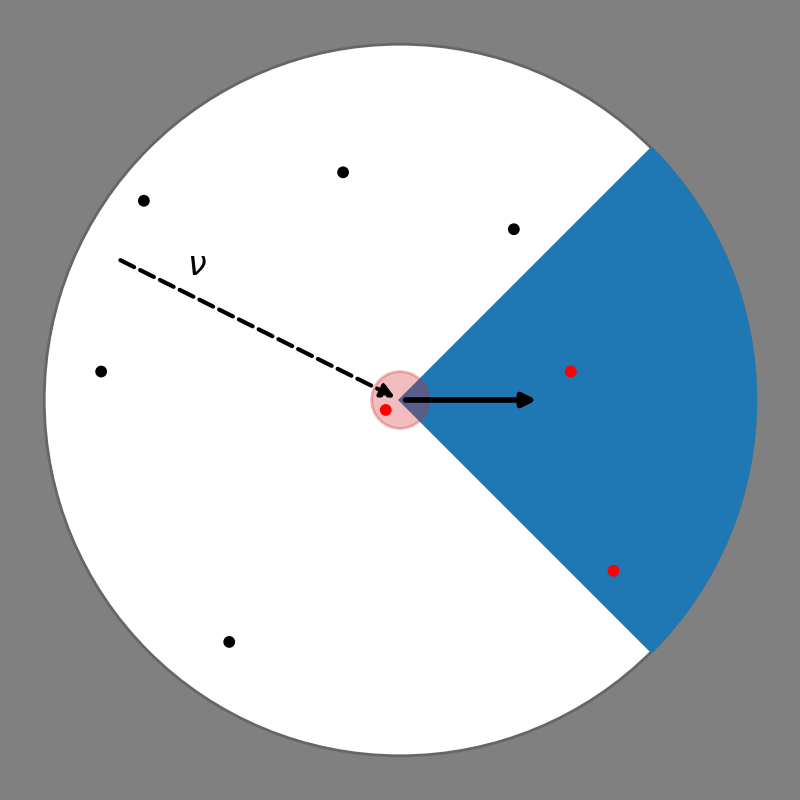}
    \includegraphics[width=0.89\linewidth]{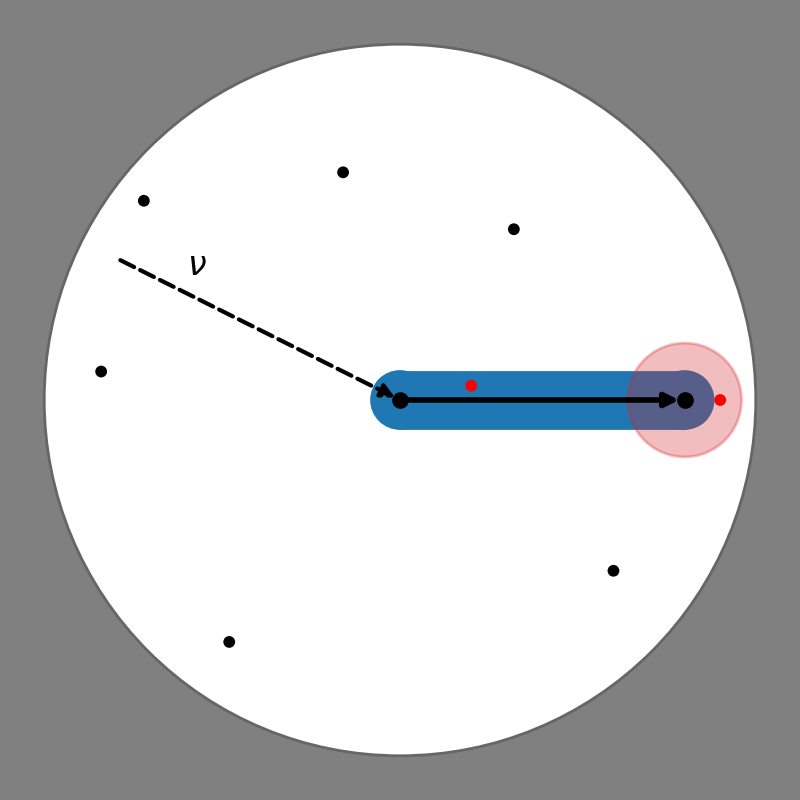}
    \caption{A cartoon illustrating applied topological selection cuts. The top figure shows the cone cut around the shower in blue, and the $\gamma$ cut around the primary vertex in pink, for $\nu_e/\bar{\nu}_e$ events. The bottom figure shows the PoCA cut around the track in blue, and the $\gamma$ cut around the end of the muon track in pink, for $\nu\mu/\bar{\nu}_\mu$ events.}
    \label{fig:cuts}
\end{figure}

To remove lepton-associated blip activity, we introduce two lepton-related topological selections, as shown in Figure~\ref{fig:cuts}.  
In $\nu_e/\bar\nu_e$ charged current interactions, we apply a cone-shaped angle cut to the electron shower, removing blips within a $45^\circ$ cone from the shower direction, as shown in the blue region in the top subfigure in Figure~\ref{fig:cuts}.
In $\nu_\mu$ and $\bar\nu_\mu$ charged current interactions, we connect the primary interaction vertex and the end point of the muon track with a line segment as an approximation to the muon track, and require the point of closest approach (PoCA~\cite{Schultz:2004kx}) distance between the blip and the line segment to be larger than \SI{20}{cm}, as shown in the blue region in the bottom subfigure in Figure~\ref{fig:cuts}.

These topological cuts can be applied to other tracks and showers in the neutrino interactions.
For the sub-GeV atmospheric neutrino sample we consider, protons and charged pions can both form tracks in addition to the muon track, while the major contribution to showers is from the primary electron in the charged current interactions.
Therefore, we apply the PoCA cut to all major kinds of particle tracks (proton, charged pion, and muons), while only applying the cone cut to the primary electron shower. 
After these cuts, lepton-associated blips account for $25\%$ ($19\%$) in $\nu_e$ ($\bar\nu_e$) charged current interactions and $11\%$ ($22\%$) in $\nu_\mu$ ($\bar\nu_\mu$) charged current interactions.  
The same lepton cut can be applied in experimental studies with reconstructed shower direction and detailed tracking.

\subsection{Backgrounds from nuclear de-excitation $\gamma$'s}
\label{subsec:gamma}

\begin{figure*}[!ht]
   \centering
         \includegraphics[width=0.45\linewidth]{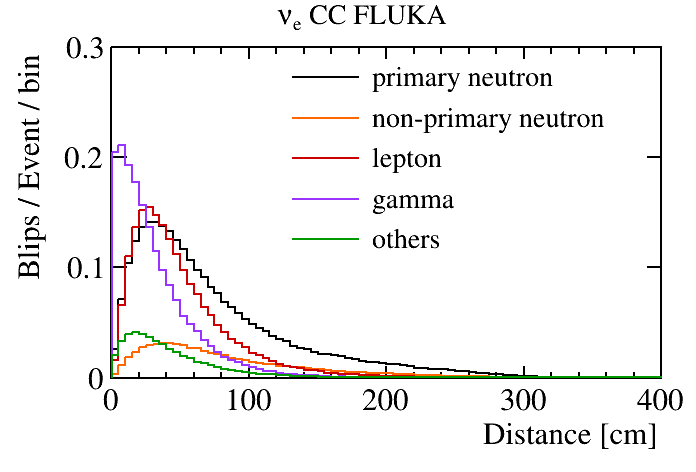}
         \includegraphics[width=0.45\linewidth]{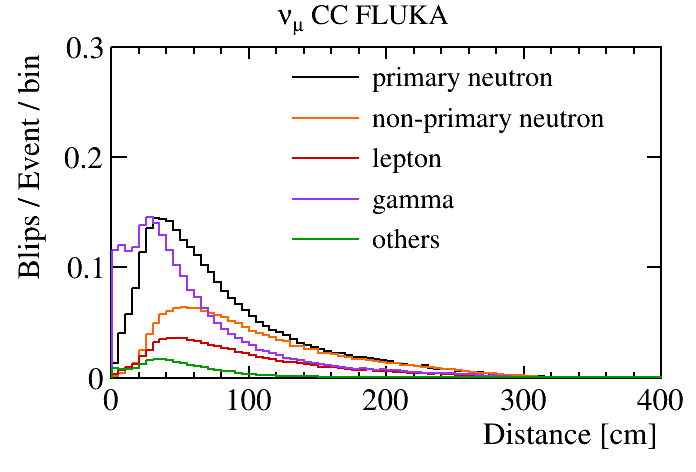}
		\\	\smallskip
         \includegraphics[width=0.45\linewidth]{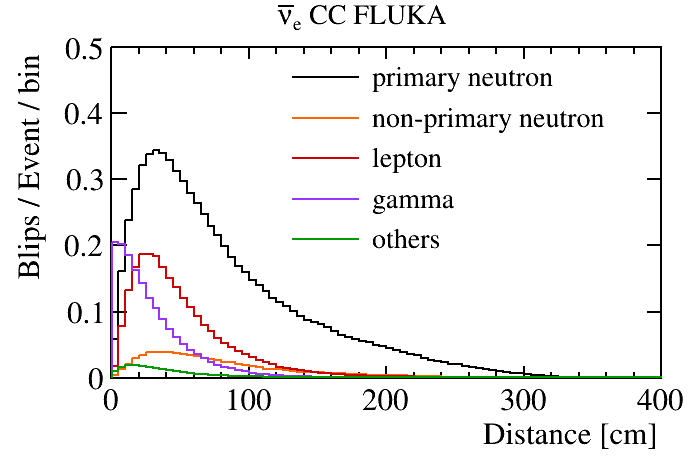}
         \includegraphics[width=0.45\linewidth]{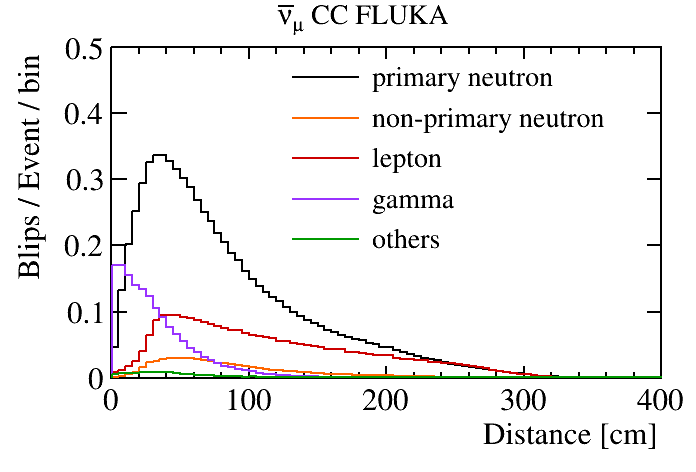}
			\caption{Distributions of distances between each blip and its primary neutrino interaction vertex for $\nu_e$ (top left), $\nu_\mu$ (top right), $\bar\nu_e$ (bottom left), and $\bar\nu_\mu$ (bottom right). The PoCA cut for $\nu_\mu/\bar\nu_\mu$'s and the cone angle cut for $\nu_e/\bar\nu_e$'s is applied, while lepton vertex and endpoint aimed at reducing de-excitation $\gamma$-ray populations are not applied.}
			\label{fig:blipD_step2}
\end{figure*}

After lepton-related PoCA or cone topological selections, the blip distribution as a function of the distance to the primary vertex is shown in Figure~\ref{fig:blipD_step2}.  
At locations close to the neutrino vertex, a majority of blips appear to have $\gamma$-related ancestry.  
The major contributor to this category is primary de-excitation $\gamma$-rays from the struck nucleus hosting the neutrino interaction.  
Blips from primary $\gamma$-rays are more densely populated near the primary vertex, since the interaction lengths of MeV-scale $\gamma$-rays are shorter than that of energetic neutrons in liquid argon.  
To reject these primary $\gamma$-rays blips, we implement a cut removing blips less than \SI{20}{cm} away from the neutrino interaction vertex.  

For the $\nu_\mu$ charged current interactions pictured in Figure~\ref{fig:blipD_step2}, one can observe a higher blip contribution from non-primary neutrons.  
In addition, $\gamma$-related blips tend to be distributed towards a larger distance range from the primary vertex than for the other three neutrino channels.  
These features reflect the presence of de-excitation products of stopped $\mu^-$-argon nuclear capture.
To remove blips from capture $\gamma$-ray products as well as Michel-electrons, we require blips to be at least \SI{40}{cm} away from the end point of muons in $\nu_\mu$ charged current interactions.  

\subsection{Backgrounds from Non-Primary Neutrons}
\label{subsec:other}

Non-primary neutrons not directly from the neutrino interactions can be produced during the propagation of final-state particles, most notably by the hadronic interaction of protons and pions with the argon nuclei, and also nuclear processes such as $\mu^-$ capture and de-excitation.  
The signatures of these neutrons are identical to primary neutrons and are indistinguishable except for their spatial distribution: secondary neutrons are more prone to be near the production vertex, which is typically different from the neutrino interaction vertex.
Due to the long interaction length of neutrons in liquid argon, it is not feasible to design a simple removal cut for these non-primary neutrons, especially for sub-GeV neutrino interactions where the spatial spread of the event is often smaller than or comparable to the neutron interaction length.  
Instead, non-primary neutrons may be statistically accounted for in certain analyses using correlated indicators, such as the existence of non-muon originated Michel $e^+$'s (indicating the potential presence of stopped $\pi^+$'s), kinked tracks (indicating inelastic nuclear scattering), or tracks featuring no Bragg peak (indicating nuclear absorption in flight).  
Since this analysis represents a simple, general demonstration of neutron reconstruction capabilities, we do not implement any specialized cuts to reduce this background category.  

\subsection{Backgrounds from Non-Neutrino Sources}
\label{subsec:nonnu}

Non-neutrino originated radiation can deposit energy in liquid argon and generate blips.
These energy depositions typically come from TPC-internal or TPC-external natural radioactivity or cosmogenic activity.  
In the on-surface MicroBooNE neutrino LArTPC experiment, it has been demonstrated that $\beta$-decays of $^{39}$Ar in the LAr bulk dominate blip production at energies $<$0.5~MeV, cosmic rays dominate at energies $>$3~MeV, and $^{238}$U/ $^{232}$Th/$^{40}$K $\beta$-decays mostly at or outside the surfaces of the LAr bulk dominate between these energy ranges~\cite{MicroBooNE:2024prh}.  
Of these three populations, the low-energy $^{39}$Ar component is the highest-rate contributor, particularly in regions removed from the TPC field cage edges.  
For cosmogenic activity or activation from cosmic-ray muon interactions, rates of blip production will be much reduced for underground sites relative to that demonstrated in MicroBooNE, and could be further reduced using time and spatial correlations with reconstructed cosmic-ray muon tracks.  
For natural radioactivity, most are introduced by materials near the detector boundary, such as field cage, photo-sensor, and electronics, and can be reduced by introducing a spatial fiducial volume cut.  
The remaining major background for neutron reconstruction is the intrinsic radioactivity from $^{39}$Ar.  

Since the beta decay spectrum of $^{39}$Ar has an end point of \SI{0.56}{MeV}~\cite{Mougeot:2015bva} and only produces true blip energies below this value, we apply a minimum blip energy cut of \SI{0.6}{MeV} to reduce this background contribution.  
In anticipation of needing to further reduce residual $^{39}$Ar contamination from mis-reconstruction and blips from other radiogenic contributors, we require all the blips used in this analysis to be contained within \SI{2.5}{m} of their corresponding neutrino interaction vertex.
The expected $^{39}$Ar blip count is $<0.1$ per neutrino interaction after application of this cut, and we neglect this background source for the remainder of our analysis.  

\subsection{Final Selected Blip Populations}

\begin{figure*}[!ht]
   \centering
         \includegraphics[width=0.45\linewidth]{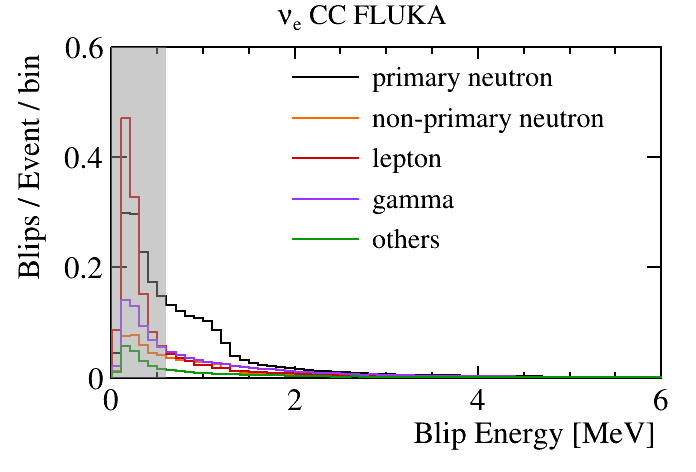}
         \includegraphics[width=0.45\linewidth]{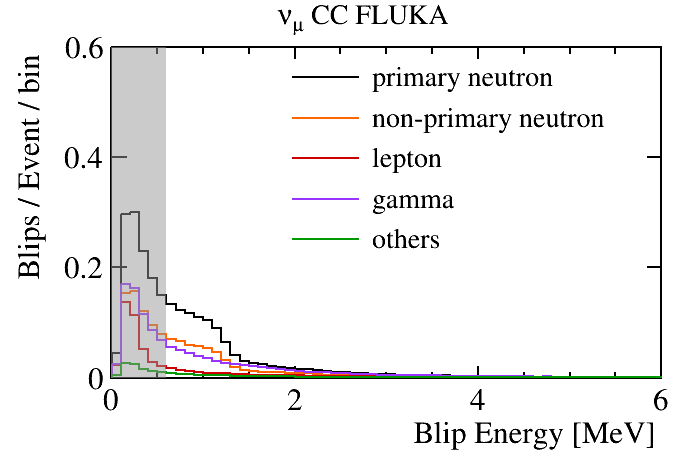}
		\\	\smallskip
         \includegraphics[width=0.45\linewidth]{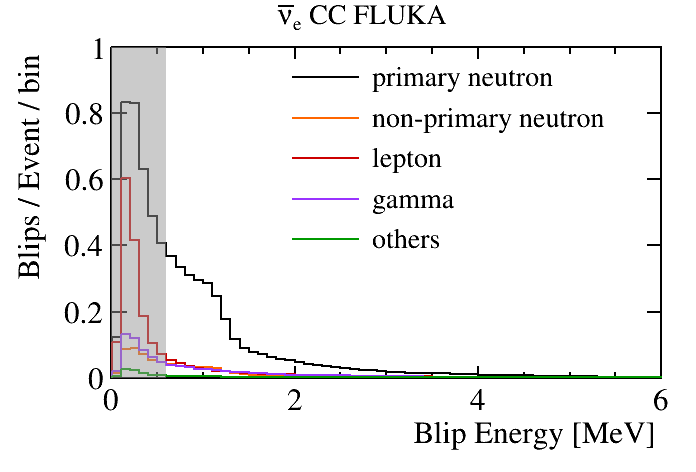}
         \includegraphics[width=0.45\linewidth]{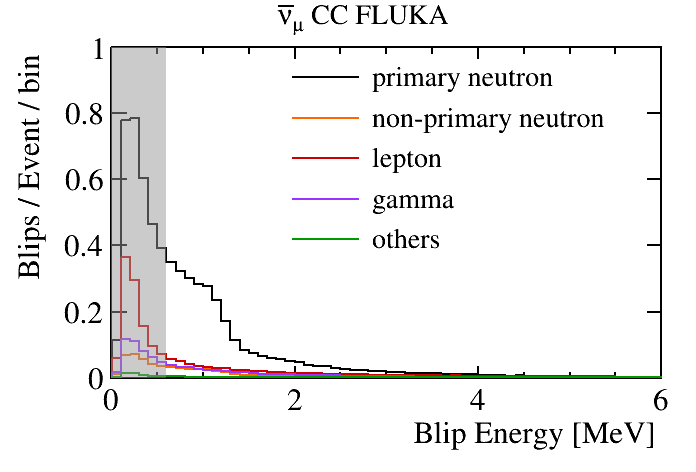}
			\caption{Reconstructed energy distributions for selected blips from $\nu_e$ (top left), $\nu_\mu$ (top right), $\bar\nu_e$ (bottom left), and $\bar\nu_\mu$ (bottom right) interactions. All topological cuts, including PoCA cut for $\nu_\mu/\bar\nu_\mu$'s and the cone angle cut for $\nu_e/\bar\nu_e$'s, the primary vertex cut, and the end-of-muon-track cut, are applied, as shown in Figure~\ref{fig:cuts}.}
			\label{fig:blipE_step3}
\end{figure*}

After applying the topological background rejection in Sections~\ref{subsec:lepton} and~\ref{subsec:gamma}, using the PoCA and cone cuts, the $\gamma$ rejection cut at the neutrino interaction vertex and the muon end point, the reconstructed blip energy spectra from neutrino interactions are shown in Figure~\ref{fig:blipE_step3}.  
For all channels, neutrons appear to be the dominant producers of blips with $>$0.5~MeV reconstructed energy, while final-state leptons contribute more substantially at the lowest blip energies.  
Neutron blip energy spectra contain a clear edge between 1.0 and 1.5~MeV, a feature consistent with Compton scattering of 1.46~MeV $\gamma$-rays emitted in transitions between the first excited state and ground state of the $^{40}$Ar nucleus.  
Meanwhile, the exponentially falling lepton blip spectra match what one would expect from  stochastic radiative energy losses from $\delta$-ray and final-state $e^+$ or $e^-$. 
This validates our \SI{0.6}{MeV} energy cut in Section~\ref{subsec:nonnu} against the $^{39}$Ar background.

After the \SI{0.6}{MeV} energy cut in addition to the topological cuts, the blip multiplicity in the final selected blip sample for each neutrino type is shown in Table~\ref{tab:finalsample}, with breakdowns in each blip type.
\begin{table}[!ht]
    \centering
    \begin{tabular}{cccccc}
		 \hline
		 \hline
		 & primary $n$ & non-primary $n$ & $l$ & $\gamma$ & other\\
		 \hline
		 $\nu_e$ & 1.07 & 0.29 & 0.38 & 0.50 & 0.15\\
		 $\bar\nu_e$ & 3.06 & 0.35 & 0.47 & 0.42 & 0.07\\
		 $\nu_\mu$ & 1.12 & 0.56 & 0.20 & 0.59 & 0.09\\
		 $\bar\nu_\mu$ & 2.98 & 0.27 & 0.75 & 0.41 & 0.04\\
       \hline
       \hline
    \end{tabular}
	 \caption{Multiplicity of the final selected blip sample for each neutrino type, with each broken down by ancestry category.}
    \label{tab:finalsample}
\end{table}

In the Supplementary Material~\cite{appendix}, we provide various associated blip distributions, including vertex distance, true energy spectrum, and reconstructed energy spectrum, before and after background rejection cuts, using both the \texttt{FLUKA} and \texttt{LArSoft}-\texttt{Geant4} particle transports. 
 
\section{Neutron Reconstruction Performance}
\label{sec:neutronrec}

Neutron blips carry information on the existence, multiplicity, and kinematics of final-state neutrons.
When multiple primary neutrons are produced in a neutrino interaction, it is highly difficult to separate blips from each of the primary neutrons due to the spread of blips from neutron free path length.
We therefore study neutron reconstruction using the entire neutron system, defined as the ensemble of all primary neutrons from a neutrino interaction, instead of individual neutrons.
The main goals of neutron reconstruction include the identification of these primary neutrons and the reconstruction of the direction and energy of the neutron system.

\subsection{Neutron Identification}
\label{subsec:neutronid}

\begin{figure*}[!ht]
   \centering
         \includegraphics[width=0.45\linewidth]{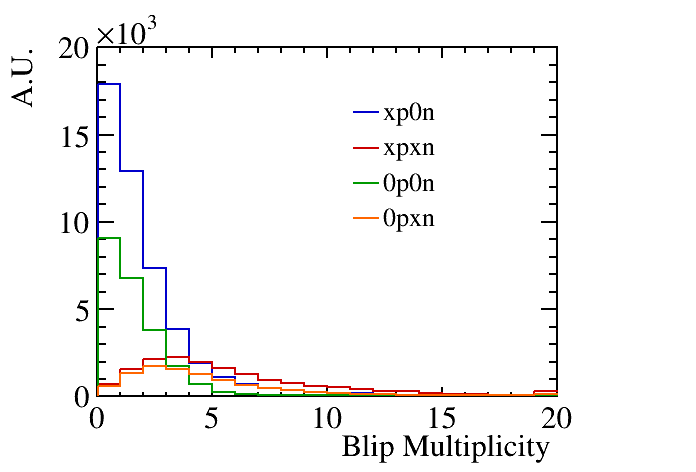}
         \includegraphics[width=0.45\linewidth]{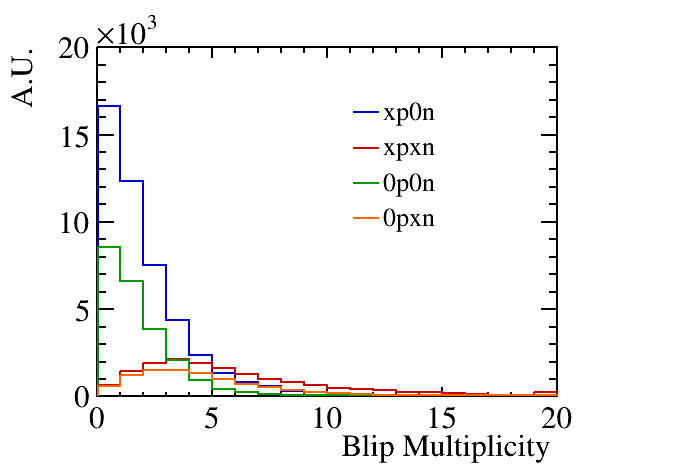}
         \includegraphics[width=0.45\linewidth]{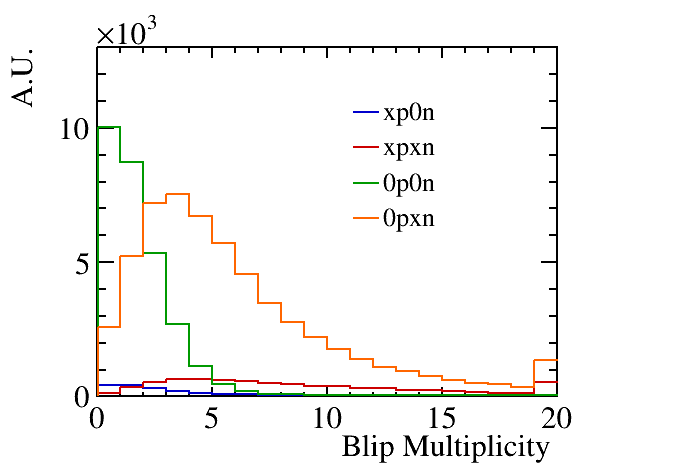}
         \includegraphics[width=0.45\linewidth]{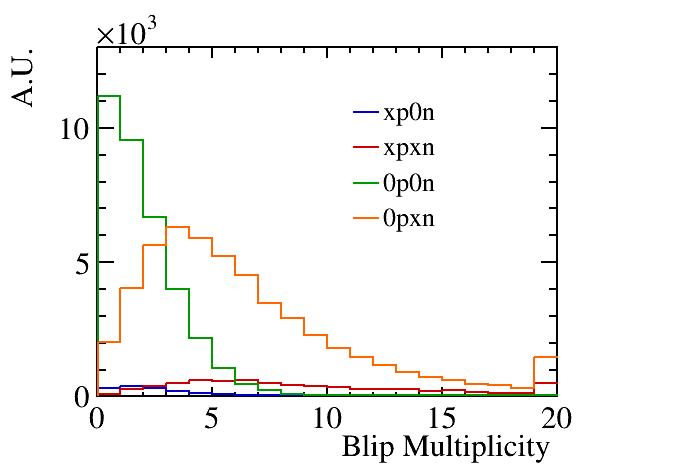}
			\caption{Blip multiplicity for different final state topologies in $\nu_e$ (top left), $\nu_\mu$ (top right), $\bar\nu_e$ (bottom left), and $\bar\nu_\mu$ (bottom right) interactions. The last bin along the x-axis serves as an overflow bin.}
			\label{fig:blipmult}
\end{figure*}

For the sub-GeV neutrinos considered in this paper, we split the simulated events in four categories based on the primary neutrino topology in the final states: with protons and no neutrons (X$p0n$), with protons and neutrons (X$p$X$n$), with no proton or neutrons ($0p0n$), and with no proton but some neutrons ($0p$X$n$).
We require the neutron categories X$n$ to have at least 1 neutron with more than \SI{5}{MeV} kinetic energy, and the proton category X$p$ to have at least 1 proton with more than \SI{30}{MeV}, for visibility.
The blip multiplicity in each category is shown in Figure~\ref{fig:blipmult}.  
In Figure~\ref{fig:blipmult} we observe an overall tendency for X$n$ samples to have more blips than $0n$ samples.
This tendency is stronger in antineutrino samples than neutrino samples, as neutrons get more energy from antineutrino interactions than neutrino interactions.

\begin{figure}[!ht]
   \centering
         \includegraphics[width=0.89\linewidth]{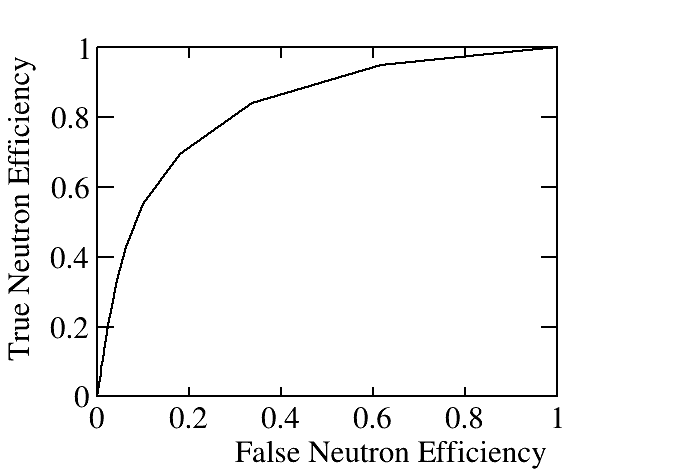}
         \includegraphics[width=0.89\linewidth]{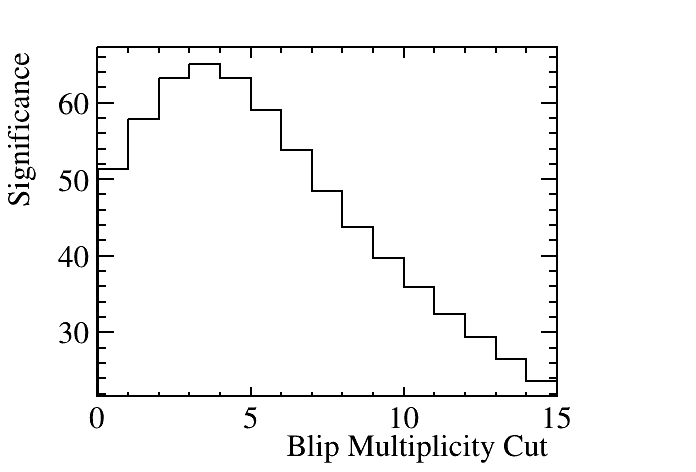}
			\caption{Top: ROC curve for the blip multiplicity cuts used for neutron identification on the $\nu_e$ sample. Bottom: Blip multiplicity distribution used for cut optimization in a mixed $\nu_e/\bar\nu_e$ sample. The best cut boundary, whose definition can be found in the text, is determined to be $>$3 blips.}
			\label{fig:nueROC}
\end{figure}

To quantify the performance of neutron identification via blips, we consider a simple cut analysis using blip multiplicity only, targeting at separating X$n$ events from $0n$ events.
For each neutrino samples of $\nu_e$, $\bar\nu_e$, $\nu_\mu$, and $\bar\nu_\mu$, we first map signal efficiency (true positive rate) for X$n$ events and background contamination (false positive rate) for $0n$ events as a function of reconstructed blip multiplicity, also known as a Receiver Operating Characteristic (ROC) curve.  
An example ROC curve for the pure $\nu_e$ sample is shown in Figure~\ref{fig:nueROC}.  
The area under the curve (AUC) listed in Table~\ref{tab:AUC} is an evaluation of the neutron identification performance, with an AUC of 1 indicating perfect, background-free X$n$ signal identification, while an AUC of 0.5 indicates no separation.  
The difference in AUC between $\nu$ and $\bar\nu$ primarily arises from the higher average summed kinetic energy of final-state neutrons for the latter category, while the reduced AUC for $\nu_\mu$ comes from the confounding presence of $\mu^-$ capture-at-rest blips in both $0n$ and X$n$ events.  

\begin{table}[!ht]
    \centering
    \begin{tabular}{ccccc}
		 \hline
		 \hline
		 & $\nu_e$ & $\nu_\mu$ & $\bar\nu_e$ & $\bar\nu_\mu$ \\\hline
		 AUC & 0.83 & 0.82 & 0.85 & 0.85\\
		 \hline
		 \hline
    \end{tabular}
	 \caption{The area under curve (AUC) for the ROC Curve of the blip multiplicity selection in neutron identification, with breakdowns in neutrino types.}
    \label{tab:AUC}
\end{table}

Next, we look to optimize blip cuts events towards the best significance of selecting $X$n events in a realistic neutrino sample.
To quantify the significance, we use the approximated Asimov significance~\cite{Kumar:2015tna},
\begin{equation}
	\sqrt{2[S-B\ln(1+S/B)]}
	\label{eq:Asimov}
\end{equation}
where $S$ denotes the number of neutrino interactions with primary neutrons in the final state, X$n$, and $B$ denotes the number of neutrino interactions without primary neutrons in the final state, 0$n$. 
Figure~\ref{fig:nueROC} shows an example of a mixed $\nu_e/\bar\nu_e$ sample, using the $\nu_e/\bar\nu_e$ ratio found in sub-GeV atmospheric neutrino fluxes.  
This process is separately done for $\nu_e/\bar\nu_e$ and $\nu_\mu/\bar\nu_\mu$ as lepton flavor identification typically performs well in LArTPCs, and the residual contamination can be considered secondary.  
The best cut is found at 3 blips for both the $\nu_e/\bar\nu_e$ sample and the $\nu_\mu/\bar\nu_\mu$ sample. 

At this optimized blip multiplicity cut value, we define events with blip multiplicity $>$ best cut to be in a neutron rich subsample (reconstructed X$n$ sample), and events with blip multiplicity $\le$ best cut to be in a neutron poor subsample (reconstructed $0n$ sample).
We can then evaluate neutron identification performance using a confusion matrix, as shown in Table~\ref{tab:neutronconf}.  
It appears that, using blip multiplicity information alone, sub-GeV neutrino events hosting final-state neutrons can be identified with $\sim$70\% efficiency, with $\sim20\%$ of neutron-free events contaminating this category due to the presence of non-primary-neutron blip activity.  
This result, while promising on its own, is likely to be improved upon with additional methods that will be discussed in Section~\ref{sec:altmeth}.  

\begin{table}[!ht]
    \centering
    \begin{tabular}{ccc}
       reco X$n$ & 0.18 & 0.71\\
       \hline
       reco 0$n$ & 0.82 & 0.29 \\
       \hline
         & true 0$n$ & true X$n$\\
    \end{tabular}
    \begin{tabular}{ccc}
       reco X$n$ & 0.21 & 0.74\\
       \hline
       reco 0$n$ & 0.79 & 0.26 \\
       \hline
         & true 0$n$ & true X$n$\\
    \end{tabular}
	 \caption{Confusion matrix for $\nu_e/\bar\nu_e$'s (left) and $\nu_\mu/\bar\nu_\mu$'s (right).}
    \label{tab:neutronconf}
\end{table}

\subsection{Direction and Energy of the Neutron System}
\label{sec:neutronedir}

In this Section, we aim to compare the true direction and energy of final-state neutron systems with comparable quantities reconstructed from blip-based information.  
This is the first quantification of this kind performed in the literature for directional reconstruction.  
While blip-based neutron kinetic energy reconstruction capabilities were investigated in Ref.~\cite{Castiglioni:2020tsu}, this is the first exploration above a few \SI{10}{s} of MeV -- an energy regime particularly relevant for GeV-scale neutrino interactions.  

For directionality, we consider the direction of the neutron system along the true momentum vector $P_n$, 
\begin{equation}
	\vec P_n=\sum\vec P_i^n,
	\label{eq:P_n}
\end{equation}
where $\vec P_i^n$ is the momentum of every primary neutron $i$ above \SI{5}{MeV} in the neutron system.  
We then assign vectors $\vec V_j^b$ from the primary interaction vertex to each blip $j$ after the background reduction cuts discussed in Section~\ref{sec:bkgred}, and take the sum of these vectors $\vec V_b$, 
\begin{equation}
	\vec V_b=\sum\vec V_j^b,
	\label{eq:V_b}
\end{equation}
as the reconstructed neutron system direction.
The direction reconstruction performance is quantified as the inner angle $\theta$ between $\vec P_n$ and $\vec V_b$, 
\begin{equation}
\theta=\arccos(\vec P_n\cdot\vec V_b/|\vec P_n||\vec V_b|).
	\label{eq:theta}
\end{equation}
Different directionality reconstruction, for example, weighting blip vectors with blip energy, are also explored and are observed to change the result slightly, but does not present any significant improvement in performance.

\begin{figure}[!ht]
   \centering
         \includegraphics[width=0.89\linewidth]{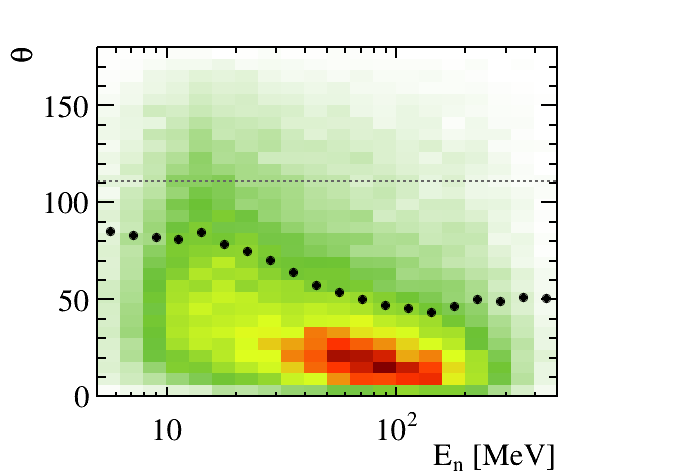}
			\caption{Direction reconstruction inner angle $\theta$ as a function of the neutron system energy, using blip information in $\bar\nu_e$ events. Black points indicate the 68\% percentile resolution from the contour. Grey dashed lines indicate the 68\% percentile resolution if no directional information is obtained.}
         \label{fig:n_perf_dir}
\end{figure}

Figure~\ref{fig:n_perf_dir} shows the direction reconstruction performance in the inner angle $\theta$ for $\bar\nu_e$ interactions as a function of the neutron system energy $E_n$, where $E_n=\sum E_n^i$, and $E_n^i$ is the kinetic energy of each neutron above \SI{5}{MeV} in the neutron system.  
Besides these $\theta-E_n$ distributions indicated by the colored contours in Figure~\ref{fig:n_perf_dir}, the 68\% percentile, or the resolution in $\theta$, is also plotted.
Similar performance is observed in neutron systems from $\bar\nu_e$, $\nu_\mu$, and $\bar\nu_\mu$ events.
At \SI{100}{MeV} kinetic energy, the direction of the neutron system can be reconstructed with $\sim$40$^\circ$ angular resolution.
While the performance is not as good as for high-energy electromagnetic showers or muon tracks, it is a meaningful outcome that demonstrates that blips can uncover the ability to perform differential cross-section measurements as a function of neutron angular kinematics for $\nu-$Ar interactions.

For energy reconstruction, we consider $E_n$, the sum of kinetic energy of each neutron above \SI{5}{MeV} in the neutron system.
The reconstruction estimator is constructed as the naive sum of the energy of the blips in MeV, $E_\text{blip}=\sum E_j$, where $E_j$ is the reconstructed energy of blip $j$ in MeV.  
We also tried different estimators, including removing the blip energy weight, or add a distance-to-vertex weight, none of which significantly outperforms the naive sum.  

\begin{figure}[!ht]
   \centering
         \includegraphics[width=0.89\linewidth]{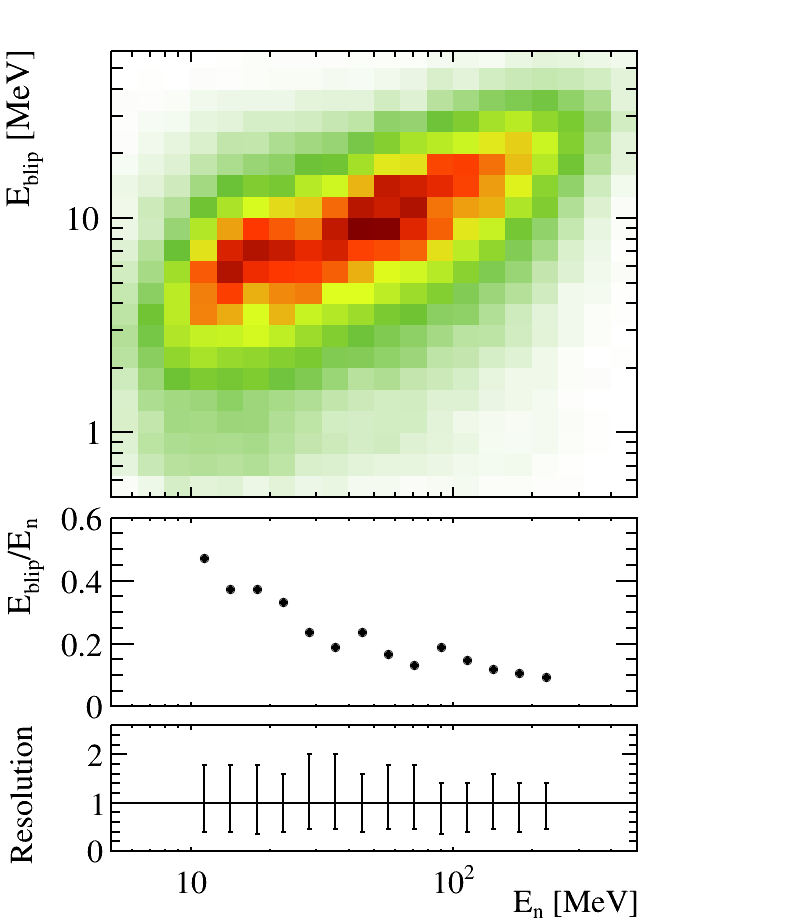}
			\caption{Energy reconstruction for the neutron system using blip information $E_\text{blip}$ as a function of true neutron system energy $E_n$ in $\bar\nu_e$ events. Top panel shows the 2-dimensional correlation. Middle panel shows the ratio of $E_\text{blip}/E_n$, where the peak in the $E_\text{blip}$ in a given $E_n$ bin is taken as the numerator, and the center of the $E_n$ bin as the denominator. Bottom panel shows the asymmetric relative resolution, defined as 68\% of population on one side of the peak in $E_\text{blip}$.}
         \label{fig:n_perf_en}
\end{figure}

Figure~\ref{fig:n_perf_en} shows the energy correlation between the sum of blip energy and the energy of the neutron system in $\bar\nu_e$'s.  
The observed correlation between true and reconstructed neutron energy indicates that blip-based neutron calorimetry is possible in LArTPC neutrino interactions, as further discussed in Section~\ref{subsec:nurec}.  
The middle panel of Figure~\ref{fig:n_perf_en} shows that across the energy range considered, the summed energy of all signal blips accounts for $\le50\%$ of the total neutron energy.  
Instead, most energy is lost in a variety of different processes: elastic neutron-argon scattering, binding energy losses from nucleon emission, un-reconstructed charge below blip reconstruction thresholds, reconstructed blips excluded from the signal sample, and neutron escape from the \SI{2.5}{m} radius.  
Energy losses are fairly evenly split between these various categories.  
The bottom panel of Figure~\ref{fig:n_perf_en} shows the fractional resolution as a function of true neutron energy.  
Above \SI{100}{MeV} kinetic energy, the energy of the neutron system can be reconstructed with $\sim$50\% energy resolution.  

A summary of the results in this Section is presented in Table~\ref{tab:neutrecsum}.
\begin{table}[!ht]
    \centering
    \begin{tabular}{cc}
		 \hline\hline
       Identification efficiency in subGeV $\nu_e/\bar\nu_e$'s & 0.71\\
       Background inefficiency in subGeV $\nu_e/\bar\nu_e$'s & 0.18\\
       Angular resolution at 100 MeV & $40^\circ$\\
       Energy resolution at 100 MeV & 50\%\\
		 Dominant uncertainties & Section~\ref{subsec:neutprop}\\
       \hline\hline
    \end{tabular}
	 \caption{Summary of demonstrated neutron identification and reconstruction performances in Section~\ref{sec:neutronrec}.}
    \label{tab:neutrecsum}
\end{table}

\section{Physics Applications}
\label{sec:physapp}

In this section, we discuss two example physics use cases of blip-based neutron reconstruction: neutrino-antineutrino separation, and neutrino kinematic reconstruction.

\subsection{Neutrino Antineutrino Separation}

\subsubsection{Separation in Atmospheric Neutrinos}

Separation of neutrinos from antineutrinos is an important topic in atmospheric neutrinos, especially for the purposes of improving CP violation and mass ordering sensitivities.
Neutron reconstruction is a typical approach towards neutrino antineutrino separation~\cite{He:2026aag}.
The recent Super-Kamiokande atmospheric study has demonstrated that with a modest 26\% efficiency in neutron identification, an improvement in the sample purity and thus the mass ordering sensitivity can be seen~\cite{Super-Kamiokande:2023ahc, Super-Kamiokande:2025cht}.

Charge conservation implies that in neutrino interactions, statistically speaking, neutrinos tend to produce more protons, and antineutrinos tend to produce more neutrons.
In LArTPCs, the high image resolution enables the separation of $\nu$'s and $\bar\nu$'s by the identification of protons in final state particles in neutrino interactions.  
In the sub-GeV atmospheric neutrino charged current interaction samples used in this analysis, assuming a perfect identification of protons above \SI{30}{MeV} energy, for electron-flavor neutrinos, one can obtain 98\% $\nu$ purity by requiring at least 1 reconstructed proton X$p$, and 28\% $\bar\nu$ purity by requiring no reconstructed protons $0p$, as shown in the top rows in Table~\ref{tab:nueconf}.
The performance is satisfactory for $\nu$'s but much less so for $\bar\nu$'s, due to both the flux composition and the cross section.

\begin{table}[!ht]
    \centering
    \begin{tabular}{ccccc}
		 \hline
		 \hline
		 topology & \multicolumn{2}{c}{efficiency} & \multicolumn{2}{c}{purity}\\
		 \hline
		 0$p$   & 0.34 & 0.90 & 0.72 & 0.28\\
		 X$p$   & 0.66 & 0.10 & 0.98 & 0.02\\
		 \hline
       0$p$X$n$ & 0.10 & 0.49 & 0.57 & 0.43\\
       0$p$0$n$ & 0.24 & 0.41 & 0.80 & 0.20\\
       X$p$X$n$ & 0.22 & 0.08 & 0.95 & 0.05\\
       X$p$0$n$ & 0.44 & 0.02 & 0.99 & 0.01\\
       \hline
       & $\nu_e$ & $\bar\nu_e$ & $\nu_e$ & $\bar\nu_e$ \\
       \hline
       \hline
    \end{tabular}
	 \caption{Efficiency (normalized per neutrino type) and purity (normalized per topology) in each reconstructed topology category assuming atmospheric $\nu_e$ and $\bar\nu_e$ charged current events composition.}
    \label{tab:nueconf}
\end{table}

The purity of the $\bar\nu$ sample can be enhanced by using neutron identification with blips.  
We look at the sub-GeV atmospheric neutrino charged current interaction samples to demonstrate this capability.
Using the blip multiplicity cuts for different flavor samples developed in Section~\ref{subsec:neutronid}, we can construct a topological classification of X$n$ and $0n$, where X$n$ denotes a reconstructed neutron rich category with blip multiplicity above the cut (3 blips for $\nu_e/\bar\nu_e$'s and 3 blips for $\nu_\mu/\bar\nu_\mu$'s, as detailed in Section~\ref{subsec:neutronid}), and $0n$ denotes a neutron poor category with blip multiplicity otherwise, in addition to the X$p$ and $0p$ categories.  
The $\nu_e/\bar\nu_e$ efficiencies in each topological category is listed in the left columns in Table~\ref{tab:nueconf}.
Placed in the context of atmospheric neutrinos, with the expected atmospheric neutrino flux composition and the corresponding $\nu_e/\bar\nu_e$ charged current interaction rate, we obtain the purity in each reconstructed topological category as shown in the right columns of Table~\ref{tab:nueconf}.
The $\bar\nu_e$ purity can be enhanced from 28\% in the $0p$ sample to 43\% in the $0p$X$n$ sample, at 49\% efficiency.
Super-Kamiokande reported in its atmospheric neutrino oscillation analysis~\cite{Super-Kamiokande:2023ahc} a $\bar\nu_e$ charge purity of 38.7\% at 43.6\% efficiency in the SK IV-V $\bar\nu_e-$like 1 $n$ sample.
Our result indicates a potential increase in both purity and efficiency compared to the reported Super-Kamiokande result. 
Moving the blip multiplicity cut up to $5$ towards better purity, it is feasible to achieve $\sim50\%$ with $30\%$ efficiency, from blip information alone.  

For $\nu_\mu/\bar\nu_\mu$ separation, further identification can be pursued via the Michel electron from muon decays. 
The SBND experiment has demonstrated a Michel electron selection with machine learning reconstruction achieving an efficiency of $\sim$80\%~\cite{Oza2025NuIntPoster}, with potential improvement using the complementary muon capture-at-rest feature~\cite{Castiglioni:2020tsu}.
While the separation performance via blip identification is similar to $\nu_e/\bar\nu_e$'s, with the requirement of reconstructed Michel electrons (denoted as 1$m$), assuming a Michel electron selection efficiency of 80\%~\cite{Oza2025NuIntPoster}, we obtain the event numbers in each reconstructed topological category as shown in Table~\ref{tab:numuconf}.
The $\bar\nu_\mu$ purity can be enhanced from 66\% in the $0p$1$m$ sample to 77\% in the $0p$X$n$1$m$ sample.

\begin{table}[!ht]
    \centering
    \begin{tabular}{ccccc}
		 \hline
		 \hline
		 topology & \multicolumn{2}{c}{efficiency} & \multicolumn{2}{c}{purity}\\
		 \hline
		 0$p$1$m$ & 0.07 & 0.72 & 0.34 & 0.66\\
		 1$p$1$m$ & 0.13 & 0.07 & 0.91 & 0.09\\
		 \hline
		 0$p$X$n$1$m$ & 0.02 & 0.40 & 0.23 & 0.77 \\
       0$p$0$n$1$m$ & 0.05 & 0.33 & 0.44 & 0.56 \\
       X$p$X$n$1$m$ & 0.05 & 0.06 & 0.82 & 0.18 \\
       X$p$0$n$1$m$ & 0.08 & 0.01 & 0.97 & 0.03 \\
		 \hline
       & $\nu_\mu$ & $\bar\nu_\mu$ & $\nu_\mu$ & $\bar\nu_\mu$ \\
       \hline
       \hline
    \end{tabular}
	 \caption{Efficiency (normalized per neutrino type) and purity (normalized per topology) in each reconstructed topology category assuming atmospheric $\nu_\mu$ and $\bar\nu_\mu$ charged current events composition, taking into consideration Michel electron reconstruction efficiency at 80\%.}
    \label{tab:numuconf}
\end{table}

\subsubsection{Separation in Beam Neutrinos}

\begin{figure}[!ht]
   \centering
         \includegraphics[width=0.89\linewidth]{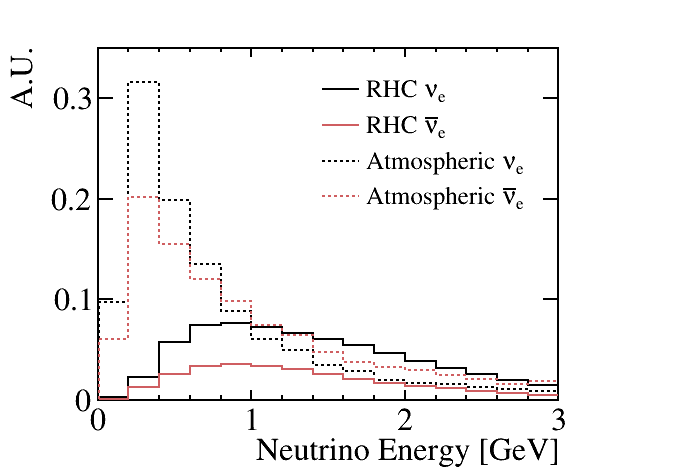}
         \includegraphics[width=0.89\linewidth]{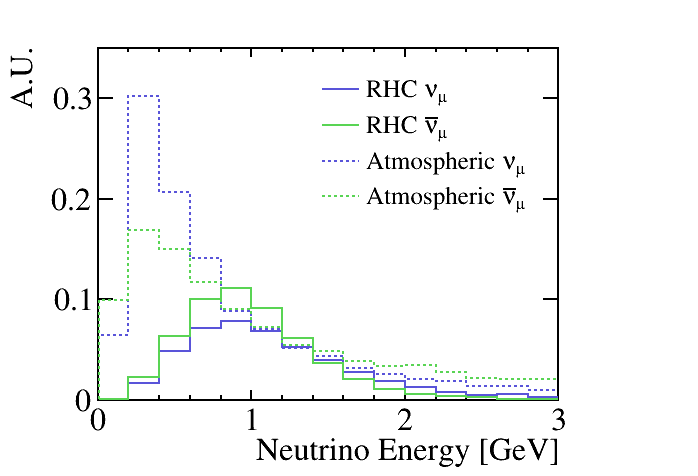}
			\caption{Energy spectra of $\nu/\bar\nu$ interactions in atmospheric neutrinos (dashed lines) and in BNB reverse-horn-current beam neutrinos~\cite{SBND:2025lha}, normalized by flavor in each neutrino source. The top figure shows $\nu_e$ (black) and $\bar\nu_e$ (red). The bottom figure shows $\nu_\mu$ (blue) and $\bar\nu_\mu$ (green).}
			\label{fig:bnbrhc_spec}
\end{figure}

The same approach can be deployed to beam neutrinos with reverse horn current (RHC), where we have the same challenge of separating $\bar\nu$'s from $\nu$'s due to the less pure $\bar\nu$ beam compared to $\nu$ beam in the forward horn current mode.
Take the Booster Neutrino Beam (BNB)~\cite{Stancu:2001cpa} as an example.
The similarity in flux and event rate energy spectra indicates applicability of our established methods, as shown in Figure~\ref{fig:bnbrhc_spec}, with the caveat of missing events above \SI{1}{GeV} neutrino energy.
Reweighting the simulated atmospheric neutrino components to that of BNB RHC~\cite{SBND:2025lha} and applying the same procedure of neutron-enhanced neutrino-antineutrino separation, we obtain an estimated $\nu_e/\bar\nu_e$ separation performance as in Table~\ref{tab:bnbnuanu}.
The $\bar\nu$ purity from BNB RHC is better compared to atmospheric neutrinos because of both the more dominant $\bar\nu$ components and the energy spectra peaking at higher energy, enabling more high energy neutrons to be detected and reconstructed via blips.
With beam neutrinos, the separation of $\nu/\bar\nu$ can include more information such as the angle between the lepton direction and the beam direction, indicating helicity suppression. 
It is typically handled using multi-variate analysis techniques, and the addition of neutron information from blip reconstruction offers a power additional differentiating variable in this kind of framework.

\begin{table}[!ht]
    \centering
    \begin{tabular}{ccccc}
		 \hline
		 \hline
		 topology & \multicolumn{2}{c}{$\nu_e/\bar\nu_e$ purity} & \multicolumn{2}{c}{$\nu_\mu/\bar\nu_\mu$ purity}\\
		 \hline
		 0$p$(1$m$) & 0.48 & 0.52 & 0.08 & 0.92\\
		 1$p$(1$m$) & 0.94 & 0.06 & 0.61 & 0.38\\
		 \hline
		 0$p$X$n$(1$m$) & 0.32 & 0.68 & 0.05 & 0.95 \\
       0$p$0$n$(1$m$) & 0.59 & 0.41 & 0.11 & 0.89 \\
       X$p$X$n$(1$m$) & 0.87 & 0.13 & 0.43 & 0.57 \\
       X$p$0$n$(1$m$) & 0.98 & 0.02 & 0.84 & 0.16 \\
		 \hline
       & $\nu_e$ & $\bar\nu_e$ & $\nu_\mu$ & $\bar\nu_\mu$ \\
       \hline
       \hline
    \end{tabular}
	 \caption{Purity (normalized per topology) in each reconstructed topology category assuming BNB RHC charged current events composition, taking into consideration Michel electron reconstruction efficiency at 80\% with 1$m$ requirement only for $\nu_\mu$'s and $\bar\nu_\mu$'s.}
    \label{tab:bnbnuanu}
\end{table}

\subsection{Neutrino Direction and Energy Reconstruction}
\label{subsec:nurec}

We typically rely on the lepton information (for example, Ref.~\cite{MicroBooNE:2019nio, MicroBooNE:2021ppm}) or leptonic energy plus information from final-state protons and mesons (for example, Ref.~\cite{MicroBooNE:2018neo, MicroBooNE:2020akw}) to represent the incoming neutrino kinetics in a LArTPC.  
For neutrino interactions producing neutrons in their final states, the reconstruction of the neutron system could in principle enhance neutrino direction and energy reconstruction by adding the missing neutron information back.
We therefore study neutrino energy and direction reconstruction using the neutron information with the technique described in Section~\ref{sec:neutronedir}.

To focus on the impact of neutron system reconstruction, we study a subsample of $\bar\nu_e$ reconstructed as X$n$, with true primary neutrons above \SI{5}{MeV} in the final state.
We first construct a toy reconstruction without neutron information as the evaluation baseline with 
\begin{equation}
\vec P_\nu^\text{rec}=|\sum\vec P_i|
	\label{eq:P_nu_rec}
\end{equation}
, where $P_i$ is the smeared momentum of each particle $i$ above the corresponding threshold using an assumed resolution.
For the sub-GeV atmospheric neutrino sample used in this analysis, we include the particles of $e$'s, $\mu$'s, $\pi^\pm$'s, and $p$'s, with corresponding kinetic energy thresholds at \SI{10}{MeV} for the leptons and \SI{30}{MeV} for the hadrons, assumed energy resolutions at 5\% for leptons and 10\% for hadrons, and assumed angular resolutions at $2^\circ$ for leptons and $10^\circ$ for hadrons.
This toy reconstruction performance is chosen to be comparable with a few phenomenological studies~\cite{Kelly:2019itm, Kelly:2023ugn} and experimental observations~\cite{MicroBooNE:2024yzp}.

\begin{figure*}[!ht]
   \centering
         \includegraphics[width=0.45\linewidth]{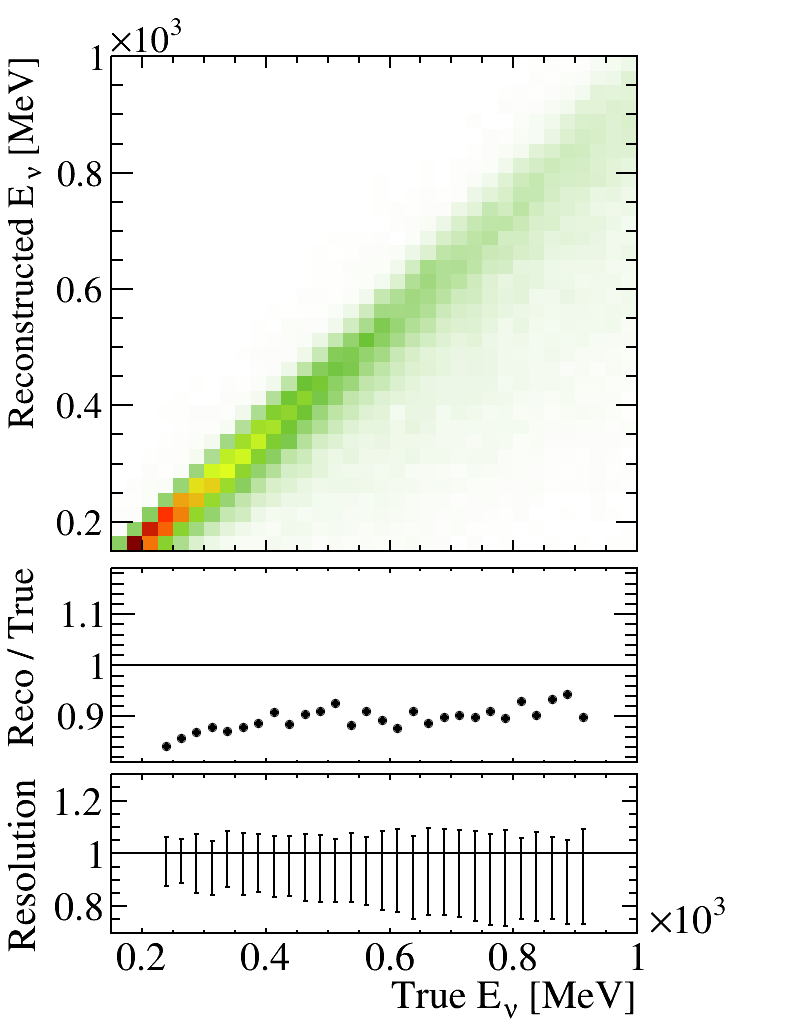}
         \includegraphics[width=0.45\linewidth]{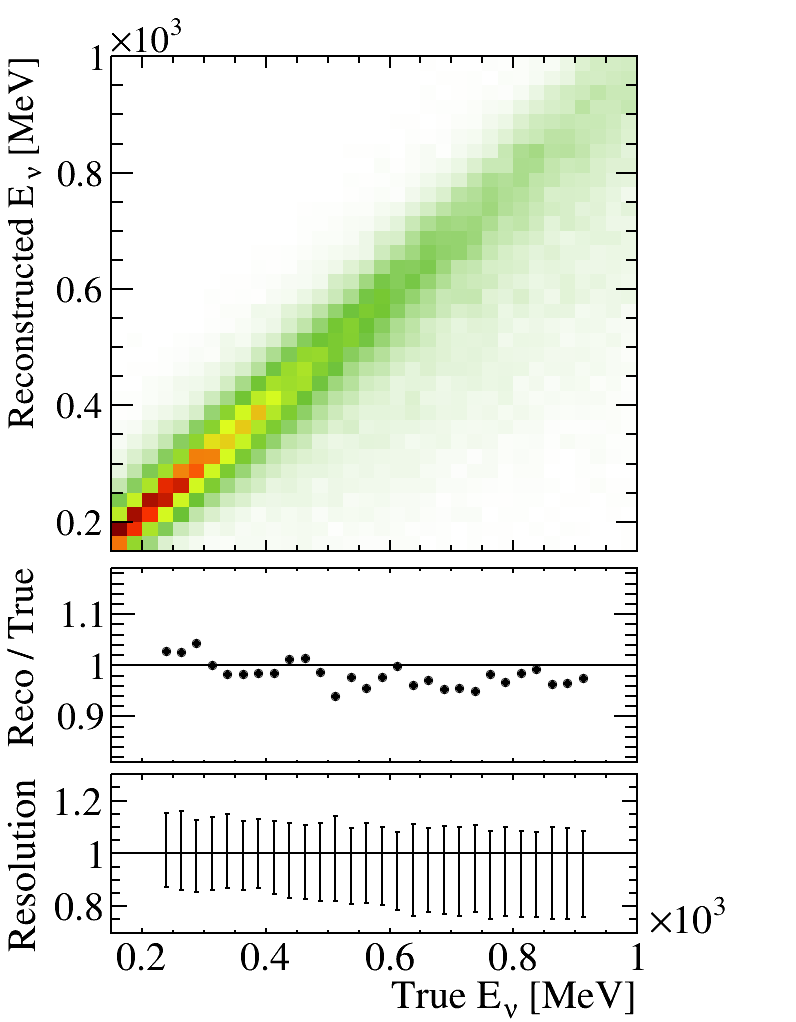}
			\caption{Energy reconstruction performance for $\bar\nu_e$ with at least 1 primary neutron above \SI{5}{MeV}, without (left) and with (right) blip information. Top: 2-dimensional response matrix relating true neutrino energy to reconstructed neutrino energy.  Middle: Ratio of reconstructed to true neutrino energy, where the peak in the reconstructed $E_\nu$ in a given true $E_\nu$ bin is taken as the numerator and the center of the true $E_\nu$ bin is taken as the denominator. Bottom: Asymmetric relative resolution defined as 68\% of the population on each side of the peak.}
			\label{fig:anue_e}
\end{figure*}

Figure~\ref{fig:n_perf_en} shows the correlation between the sum of blip energy and the energy of the neutron system.
It can be fit by quadratic functions
\begin{equation}
	E_n^\text{rec}=\text{\SI{27.8}{MeV}}+3.41\times E_\text{blip}-\text{\SI{0.0164}{MeV^{-1}}}\times E_\text{blip}^2,
	\label{eq:E_n_e}
\end{equation}
for $\nu_e/\bar\nu_e's$, and 
\begin{equation}
	E_n^\text{rec}=\text{\SI{25.4}{MeV}}+4.44\times E_\text{blip}-\text{\SI{0.0045}{MeV^{-1}}}\times E_\text{blip}^2,
	\label{eq:E_n_mu}
\end{equation}
for $\nu_\mu/\bar\nu_\mu's$, where $E_n^\text{rec}$ represents the reconstructed energy of the neutron system.
The reconstructed neutrino energy is therefore assigned as 
\begin{equation}
	E_\nu^\text{rec, n}=|\vec P_\nu^\text{rec}|+E_n^\text{rec}, 
	\label{eq:E_nu}
\end{equation}
requiring reconstructed neutron energy $E_n^\text{rec}>0$.
Figure~\ref{fig:anue_e} shows the energy reconstruction performance with and without the blip-based neutron reconstruction as a function of the incident neutrino energy $E_\nu$.
With the addition of neutron information, we see a smaller bias in neutrino energy, as well as a larger resolution.  
The bias reduction comes from neutron kinematics completing the energy deposition from neutrino interactions in LArTPCs, as noted or quantified previously in the literature in both higher-energy (4~GeV $\nu_e$/$\bar{\nu}_e$~\cite{Friedland:2018vry}) and lower-energy ($<$100~MeV $\nu_e$~\cite{Castiglioni:2020tsu,DUNE:2020zfm,Gardiner:2020ulp}) neutrino contexts.  
The worse resolution is driven by the large uncertainty in the energy of the neutron system, as seen in the width of Figure~\ref{fig:n_perf_en}.
It indicates that the energy scale uncertainty from missing neutrons is smaller than the neutron energy resolution currently achieved. 

\begin{figure}[!ht]
   \centering
         \includegraphics[width=0.89\linewidth]{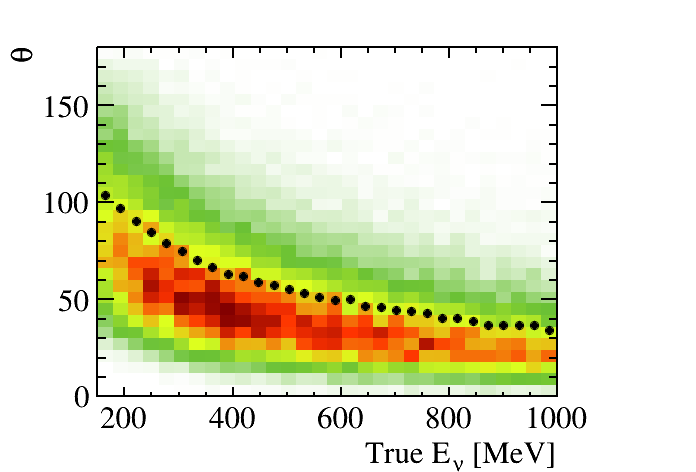}
         \includegraphics[width=0.89\linewidth]{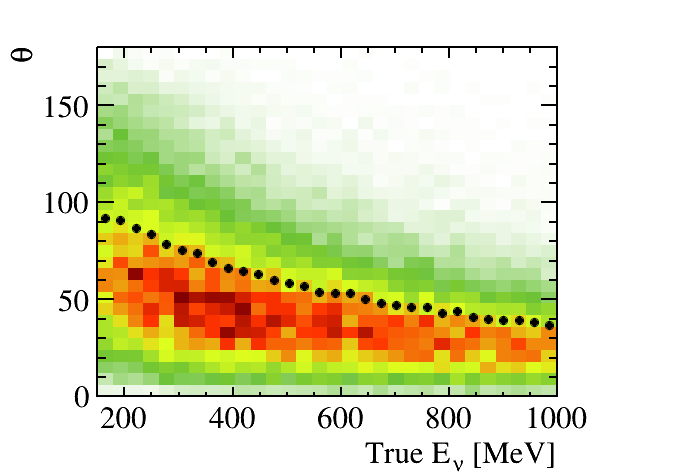}
			\caption{Direction reconstruction for $\bar\nu_e$'s with at least 1 primary neutron above \SI{5}{MeV}, without (top) and with (bottom) blip information. Black points indicate the 68\% percentile, or the resolution.}
			\label{fig:anue_dir}
\end{figure}

On directionality, the performance is evaluated using the inner angle $\theta$ between the true neutrino direction and the reconstructed neutrino direction along the momentum direction.
The neutrino direction with neutron information is constructed along the direction of the momentum with neutron information, 
\begin{equation}
	\vec P^\text{rec, n} = \vec P^\text{rec} + \vec P^\text{rec}_n,
	\label{eq:P_rec_n}
\end{equation}
	where we define $\vec P^\text{rec}_n$ as a vector with the direction of the reconstructed direction of the neutron system $\vec V_b/|\vec V_b|$, and the magnitude of $|\vec P^\text{rec}_n| = \sqrt{{E^\text{rec}_n}^2-M_n^2}$, in which $M_n$ is the neutron mass, and $E^\text{rec}_n$ is the reconstructed energy of the neutron system, requiring $E_n^\text{rec}>0$.
This definition implies an assumption of neutron multiplicity being 1, although the neutron multiplicity in the neutron system is unknown. 
The choice is justified as in a large number of events, there is one leading energetic neutron generating the majority of blips.
Figure~\ref{fig:anue_dir} shows the direction reconstruction performance with and without blip-based neutron reconstruction as a function of neutrino energy $E_\nu$.
The modest enhancement from neutron information in directional reconstruction can be seen in the lower end of the neutrino energy spectrum below \SI{300}{MeV} where the angular resolution is improved from $\sim100^\circ$ to $\sim90^\circ$.

In both the neutrino energy and neutrino direction reconstruction, we see that including blip information gives a composite result of a smaller bias and a larger spread, the former indicating the importance of including the neutron information, and the latter indicating the limited precision in our simple neutron reconstruction scheme.
One can expect potential future improvements in neutron reconstruction as discussed in Section~\ref{sec:altmeth} to enhance the precision and further benefit neutrino kinetic reconstruction.
With the current performance, however, the most robust use of blip information is likely in event classification and sample categorization through neutron identification.
As shown the left figure in Figure~\ref{fig:anue_e}, the X$n$ (neutron-rich) sample exhibits a $\sim10\%$ energy scale bias.
Thus, even in the cases which do not require neutrino-antineutrino separation, such as the highly $\nu$-pure forward-horn current beam samples, neutron identification via blips can be used to divide the data into neutron-poor (small energy-scale uncertainty) and neutron-rich (large energy-scale uncertainty) subsamples.
Dedicated neutron kinetic reconstruction will be especially valuable in physics cases where hadronic activity is directly relevant, such as hadrophilic dark matter searches~\cite{Batell:2018fqo, Ema:2020ulo, Batell:2021snh}.
In addition, the neutron kinetic reconstruction could serve as a nuisance-constraining variable in oscillation analyses~\cite{Super-Kamiokande:2023ahc}.

It should be noted that, while we demonstrate potential benefits of neutron reconstruction with blips, the resulting performance gain in neutrino physics analyses, such as oscillation studies and rare event searches, is subject to significant systematic uncertainty in neutron production modeling, as discussed in Section~\ref{subsec:neutprod}.

\section{Discussion and Outlook}
\label{sec:discuss}

\subsection{Systematic Uncertainties}

\subsubsection{Neutron Propagation and Interaction}
\label{subsec:neutprop}

Different energy depositions and blip multiplicities will originate from different modeling of neutron propagation, inelastic scattering, and argon nucleus de-excitation.  
As discussed in Section~\ref{subsec:partran}, after the neutrino interaction generation using \texttt{FLUKA}, we have the choice to propagate the produced final state particles using different packages.
While this paper adopts \texttt{FLUKA} for particle transport as the central value sample used in the studies presented in Section~\ref{sec:bkgred},~\ref{sec:neutronrec},~and~\ref{sec:physapp}, we also deployed a \texttt{LArSoft}-based \texttt{Geant4} particle transport using the Bertini Cascade model, with high-precision modeling for low energy neutrons using the \texttt{NeutronHP} library~\cite{Plompen:2020due}.

We investigated the outcomes of applying the same set of blip background reduction cuts and neutron reconstruction algorithm developed with \texttt{FLUKA} propagated sample to the \texttt{Geant4} propagated sample, as a systematic uncertainty evaluation process.
To illustrate, Table~\ref{tab:G4syst} shows the neutron identification confusion matrix previously discussed for \texttt{FLUKA} in Section~\ref{subsec:neutronid} and Table~\ref{tab:neutronconf}, with \texttt{Geant4} and \texttt{FLUKA} results shown side by side.  
The true neutron efficiency with primary neutrons in both flavor samples are quite comparable between \texttt{Geant4} and \texttt{FLUKA} particle transport, as is the rejection inefficiency for no primary neutrons in electron-flavor final states, with $\le5\%$ relative differences.  
For muon-flavored final-states, a more substantial difference, 28\% versus 21\%, is likely due to inaccurate \texttt{Geant4} modeling of de-excitation products of the $\mu^-$ capture at rest process, as noted by LArIAT experiment~\cite{LArIAT:2024otd}.  
\begin{table}[!ht]
    \centering
    \begin{tabular}{cccc}
       \hline
       \hline
		 &	 & neutron eff. & non-neutron ineff.\\
       \hline
		 $\nu_e/\bar\nu_e$ & \texttt{FLUKA} & 0.71 & 0.18\\
		 & \texttt{Geant4} & 0.71 & 0.19\\
       \hline
		 $\nu_\mu/\bar\nu_\mu$ & \texttt{FLUKA} & 0.74 & 0.21\\
		 & \texttt{Geant4} & 0.75 & 0.28\\
       \hline
       \hline
    \end{tabular}
	 \caption{Comparison of neutron identification efficiency for $\nu_e/\bar\nu_e$'s (top) and $\nu_\mu/\bar\nu_\mu$'s (bottom).}
    \label{tab:G4syst}
\end{table}

\subsubsection{Neutron Production}
\label{subsec:neutprod}

\begin{figure*}[!ht]
   \centering
         \includegraphics[width=0.45\linewidth]{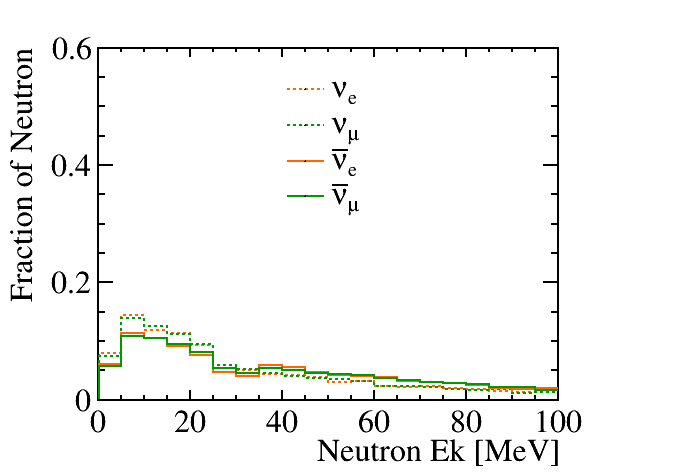}
         \includegraphics[width=0.45\linewidth]{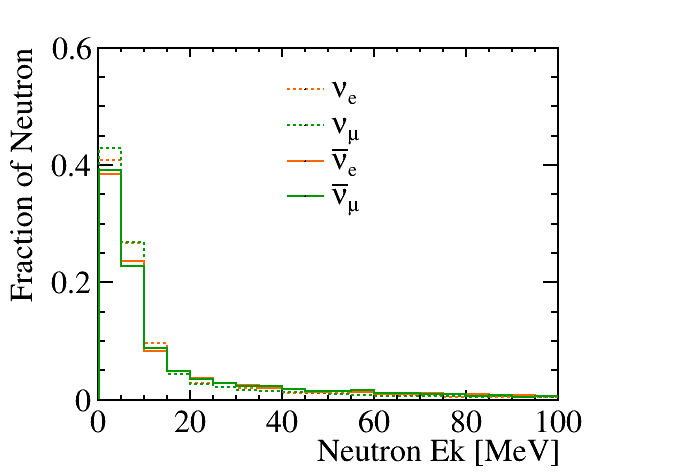}
         \includegraphics[width=0.45\linewidth]{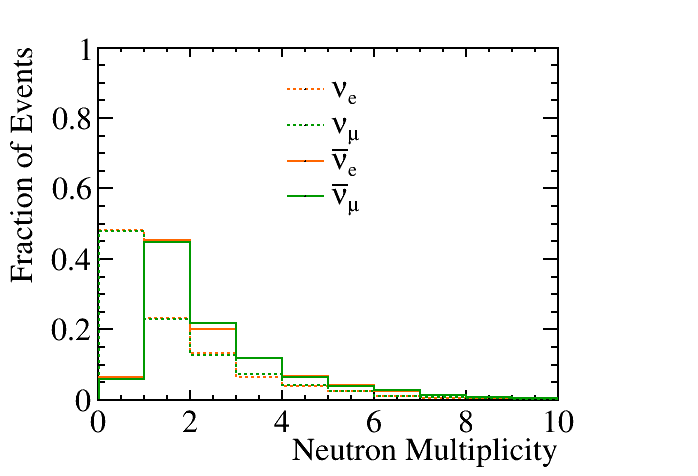}
         \includegraphics[width=0.45\linewidth]{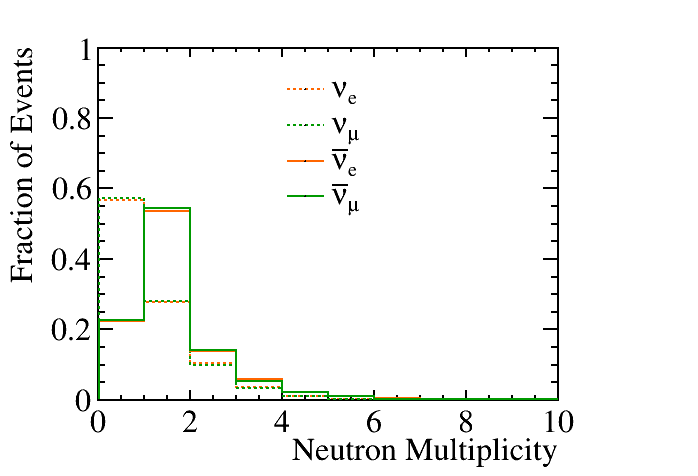}
			\caption{Neutron kinetic energy distribution with a \SI{2}{MeV} threshold (top) and neutron multiplicity distribution with a \SI{5}{MeV} threshold (bottom) from sub-GeV atmospheric neutrinos generated using \texttt{GENIE}-hA (left) and \texttt{GENIE}-INCL (right).}
			\label{fig:sys_nmultke}
\end{figure*}

It is challenging to model neutron and proton production from neutrino interaction on argon, due to the complexity of the argon nucleus.
To show the differences in the neutrino interaction modeling, we use an alternative neutrino generator \texttt{GENIE}~\cite{Andreopoulos:2009rq} in addition to \texttt{FLUKA} generator, and compare the neutron distributions from the generator predictions using the same neutrino flux.
We also consider \texttt{GENIE} with the INTRANUKE effective intranuclear rescattering model hA~\cite{GENIE:2021zuu}, and Liege Intranuclear Cascade INCL model~\cite{Liu:2026wlw}, as final state interaction variations.  
The \texttt{FLUKA} generator predictions are shown in Figure~\ref{fig:fluka_neutron}, while the \texttt{GENIE} generator predictions are shown in Figure~\ref{fig:sys_nmultke}.

The top row in Figure~\ref{fig:sys_nmultke} shows the resulting differences in neutron kinetic energy distributions for sub-GeV atmospheric neutrinos, with a \SI{2}{MeV} threshold to avoid the overwhelmingly large population of mostly-invisible evaporation neutrons (the same treatment as in Figure~\ref{fig:fluka_neutron}).  
There is a strong model discrepancy in low energy neutron production below \SI{5}{MeV} kinetic energy, which motivates our signal definition with the \SI{5}{MeV} threshold in Section~\ref{subsec:neutronid}, in addition to visibility considerations. 

The bottom row in Figure~\ref{fig:sys_nmultke} shows the neutron multiplicity above \SI{5}{MeV} kinetic energy cut.
We observe the general tendency of $\bar\nu$'s producing more neutrons than $\nu$'s as expected.
In addition, \texttt{GENIE}-INCL tends to predict a lower multiplicity than from \texttt{GENIE}-hA even with the \SI{5}{MeV} kinetic energy threshold.
This is due to \texttt{GENIE}-INCL having a similar neutron spectrum and multiplicity to \texttt{FLUKA} shown in Figure~\ref{fig:fluka_neutron}, leaning towards lower energy compared to \texttt{GENIE}-hA.
While we can raise the neutron energy threshold further to mitigate this discrepancy in the predictions, in order to use neutron information in studies such as neutrino oscillation and cross-section analyses, the reconstruction should aim to cover as much phase space as possible.  
The uncertainty in low energy neutron production therefore poses a significant challenge to use neutron information in neutrino reconstruction.
It requires more experimental data on nucleon multiplicity in LArTPC experiments, and more data-to-model comparison as well as model parameter tuning, calling for a collaborative effort from neutrino experiments, neutrino cross-section model developments, and nuclear physics experts.  

\subsection{LArTPC Detector Applicability}

The blip threshold and energy resolution, or more generally, the blip detector response, is impacted by the drift distance, wire pitch and noise environment. 
We discuss below the impact of detector parameters on our neutron blip selection and analyses:
\begin{itemize}
	\item \underline{Drift loss}: 
		Due to the impurity in LAr, the ionization electrons suffer a loss during the drift to the anode.
		This effect can be characterised as 
		\begin{equation}
			Q_a=Q_0e^{-t/\tau_e},
		\end{equation}
		where $Q_a$ is the detected charge, $Q_0$ is the deposited charge, $t$ is the drift time from blip to the anode, and $\tau_e$ is the so-called electron lifetime, a parameter indicating the purity of LAr.
		If $t$, or equivalently the blip distance to anode along drift direction (usually called $x-$coordinate), is unknown, the charge loss can result in an uncertainty during reconstruction from $Q_a$ to $Q_0$.
		However, for blips originating from neutrino interactions, such as the neutron blips and the backgrounds discussed in Section~\ref{subsec:lepton}, \ref{subsec:gamma}, and~\ref{subsec:other}, $t$ can be reconstructed with good precision assuming the blip is simultaneous (within $\sim$\SI{1}{us}) with the neutrino interaction, and thus our selection and analyses are not subject to this uncertainty. 
		With known $t$, the charge loss induces a very mild impact on energy response smearing~\cite{MicroBooNE:2023sxs} and blip threshold due to the loss of ionization electron statistics, the latter of which is further mitigated by the \SI{0.6}{MeV} blip energy cut for $^{39}$Ar background rejection, as discussed in Section~\ref{subsec:nonnu}.
	\item \underline{Signal-to-noise ratio}: 
		The signal-to-noise ratio observed in the TPC waveforms limits the blip reconstruction threshold.
		The \SI{0.6}{MeV} blip energy cut for $^{39}$Ar background rejection discussed in Section~\ref{subsec:nonnu} makes our selection not sensitive to the intrinsic signal-to-noise ratio limited blip reconstruction threshold.
		This is true for LArTPC detectors such as MicroBooNE, SBND and ProtoDUNE, with a signal-to-noise ratio of $\gtrsim20$ on the collection plane in raw data~\cite{MicroBooNE:2017qiu, Oh:2025spm, DUNE:2020cqd}.
		For future detectors such as the DUNE far detectors, maintaining a similar signal-to-noise ratio should be sufficient for neutron blip purposes.
		We note that when the signal-to-noise ratio of a LArTPC is too low, the blip energy threshold could become noise-limited rather than background-limited. 
		As an example, it will be a challenge to achieve blip-enhanced neutron reconstruction and physics application in warm electronics detectors such as ICARUS, which has a relatively low signal-to-noise ratio of 10.8 on the collection plane~\cite{ICARUS:2024hmk} due to noise from the long distance analog signal transmission.
	\item \underline{Wire pitch}: 
		Wire pitch impacts the charge deposition per wire, and thus impacts the blip reconstruction threshold.
		LArTPCs with a \SI{5}{mm} pitch, such as ProtoDUNE~\cite{DUNE:2020cqd} and DUNE~\cite{DUNE:2020ypp}, collect more charge per wire, and could use a less stringent requirement on signal-to-noise to achieve the same blip reconstruction threshold, compared to the \SI{3}{mm} pitch LArTPCs such as MicroBooNE~\cite{MicroBooNE:2017qiu} and SBND~\cite{SBND:2025lha}.
		However, as mentioned above, with the \SI{0.6}{MeV} blip energy cut for $^{39}$Ar background rejection, the blip selection in this paper is not sensitive to the intrinsic blip reconstruction threshold.
		The change of position precision between a \SI{3}{mm} pitch and a \SI{5}{mm} pitch is negligible compared to the \SI{20}{cm} and \SI{40}{cm} selection cuts in Section~\ref{subsec:lepton} and \ref{subsec:gamma}.

\end{itemize}

In addition to detector response, the background rate also limits the level information extractable from blips.
As discussed in Section~\ref{subsec:nonnu}, we consider radioactivities commonly existing in LAr bulk region while assuming the radioactivities from TPC structures are removable from a fiducial volume cut.
Cosmogenic backgrounds are not considered in this work. 
For deep-underground detectors, a generous rejection cut near the detected cosmic-ray muon track in time and space can reject most of the cosmogenic backgrounds.
For surface detectors or near-surface detectors subject to significant cosmogenic backgrounds, spallation must be studied in further detail to balance the acceptance and the background rejection efficiency, as demonstrated in other neutrino detectors~\cite{DayaBay:2017txw, DayaBay:2024xye, DoubleChooz:2018kvj, RENO:2022xbr}.

\subsection{Neutron Reconstruction Improvements}
\label{sec:altmeth}

In this paper, we demonstrated a working framework for neutron reconstruction via blips in LArTPCs using a simple and intentionally conservative approach.  
The current cuts are set to keep the performance robust when applied to various LArTPC performances.
To name a few, the \SI{20}{cm} PoCA cut is robust against the $\lesssim$\SI{1}{cm} vertex resolution in this energy range~\cite{DUNE:2026yly}.
The $45^\circ$ cone cut is robust against the $\lesssim10^\circ$ shower angular resolution in this energy range~\cite{DUNE:2026yly}, given the $\lesssim30^\circ$ intrinsic visible opening angle from the Moliere radius~\cite{ParticleDataGroup:2024cfk}.

Looking forward from this first demonstration, we see ample room for improvement in performance along many stages of the analysis process in experimental application, with the detector and reconstruction performance of a specific detector.  
Here we list some clear areas of future improvement: 

\begin{itemize}
	\item \underline{PoCA cuts for $\mu^{\pm}$:} The current PoCA cut applied along a line segment connecting the start- and end-point of a track could miss rejection of background blips if the track has been scattered, which can be recovered by a more realistic track reconstruction, removing blips within PoCA distance of all hit clustered in any track, not necessarily along a line segment. 
\item \underline{Shower cuts for $e^{\pm}$:} Shower exclusion is also highly simplified: a selection based on a more informed approach adapted from established electromagnetic calorimetric techniques, such as the Moliere radius~\cite{ParticleDataGroup:2024cfk}, could potentially provide better treatment of this blip category.  
\item \underline{Radiogenic reductions:} To reject the dominant non-neutrino induced blips from $^{39}$Ar decays, we removed all low-energy blips from this study, which substantially reduces blip statistics.  One can imagine a broad array of potential analysis improvements using topology-based discrimination.
\item \underline{Blip proximity to features of interest:} As mentioned in Section~\ref{subsec:other}, beyond the muon PoCA cut, shower cone cut, and neutrino vertex and muon end-point cuts, no effort was made to reject non-signal blips featuring correlations with high-energy event features, such as kinked tracks, Michel $e^-$ stubs, tracks missing a Bragg peak, and so on.  
\item \underline{Secondary protons:} Our analysis provides no special consideration of $p$-generated blips.  
While they are produced far more rarely than $\gamma$-induced neutron blips, they are a far more unambiguous indication of the existence of an energetic neutron.  Proper leveraging of $p$-induced blips may be particularly helpful for higher energy neutrino interactions featuring more energetic neutrons.  
\item \underline{Light-based neutron reconstruction:} Our analysis does not consider any information provided by a LArTPC's scintillation light collection sub-system.  Light-based LArTPC information offers the potential benefit of fast timing, which may be helpful in better excluding out-of-time radiogenic energy depositions or in establishing neutron kinematics using orthogonal time-of-flight techniques.  
\end{itemize}

Future improvements addressing these points will be likely to achieve substantial enhancement in neutron reconstruction capabilities beyond what we demonstrate in our study.  
In particular, application of AI/ML techniques to the problem of neutron reconstruction is well motivated in the context of the multi-dimensional blip attribute phase space provided by a LArTPC image.  
In addition to potentially enhancing blip response in signal processing~\cite{Yu:2020wxu,ArgoNeuT:2021xtd,Bhat:2025wam}, AI/ML techniques can more fully leverage correlations in inter-blip spacing, energy, location with respect to other key event topology features, and more.  

\section{Summary}
\label{sec:summary}

We construct a simulation-based study of neutron reconstruction using blip information in a generic LArTPC detector.  
Using realistic blip response from published experimental results, we demonstrate that blip counts from sub-GeV neutrino interactions with primary neutron production are statistically significantly higher than those from neutrino interactions without primary neutrons, and that these blips can be used to reconstruct attributes of these primary neutrons.  
With a straightforward and simplistic signal blip selection and a neutron identification scheme based solely on signal blip counting, we demonstrate identification of neutrons at $\sim70\%$ efficiency with $\sim20\%$ inefficiency caused by blip contributions from non-primary neutrons or non-neutron contaminations.  
Although individual neutrons in the same neutrino interaction cannot be identified separately, the energy and direction of such a neutron system can be reconstructed at $\sim$50\% energy resolution and $\sim40^\circ$ angular resolution at $\sim100$~MeV. 
We then apply these capabilities in the context of reconstructing attributes of sub-GeV neutrino interactions in a LArTPC.  
We find that in the context of atmospheric neutrinos and BNB reverse-horn-current (anti-neutrino mode) beam neutrinos, the addition of simple blip-based neutron information can potentially improve neutrino-antineutrino separation capability, which can provide potential new handles to neutrino oscillation studies and neutrino cross-section constraints, particularly for atmospheric $\bar{\nu}_e$'s. 
Demonstrations in this paper should be seen as a jumping-off point in a long-term project of neutron reconstruction improvements in LArTPCs, as opposed to a demonstration of maximal capabilities.  
AI-informed reconstruction techniques capable of identifying and exploiting multi-dimensional correlations are well-matched to the diverse and multifaceted nature of low-energy activity in LArTPC images.  

Blip features are subject to systematic uncertainties due to the insufficient modeling and experimental data on neutron-argon interactions as well as neutron production in neutrino interactions.
With different neutron propagation modeling in \texttt{FLUKA} and \texttt{Geant4}, we note $\sim5$\% differences in neutron identification efficiency and $\sim25\%$ differences in background contamination.  
Thus, further application of neutron reconstruction to neutrino physics deliverables faces challenges from neutrino interaction modeling, neutron production uncertainties, hadronic interaction modeling, and nuclear de-excitation modeling.  
Since these uncertainties are more pronounced at lower energy and higher neutron multiplicity, these uncertainties can be mitigated by treating neutrons as a system without counting individual neutrons and by requiring energy thresholds aimed at excluding consideration of low-energy neutronic final states.  
However, these limitations should be best resolved via dedicated efforts toward experimental measurements and subsequent neutron model tuning.  

\section*{Acknowledgments}

We acknowledge and thank Diego Andrade Aldana for helpful discussions regarding reconstructed blip responses in MicroBooNE, Will Foreman for assistance in using and interpreting $\texttt{BlipReco}$ outputs, Ivan Martinez Soler and Pedro A.N. Machado for helpful discussions regarding applications to neutrino oscillation, Liang Liu for helpful discussions on \texttt{GENIE}-INCL.  This work was produced by Fermi Forward Discovery Group, LLC under Contract No. 89243024CSC000002 with the U.S. Department of Energy, Office of Science, Office of High Energy Physics, at Illinois Tech under the U.S. Department of Energy, Office of Science, Office of High Energy Physics, Contract No. DE-SC0008347, and at University of Minnesota under the U.S. Department of Energy, Office of Science, Office of High Energy Physics, Contract No. DE-SC0012069.  
Publisher acknowledges the U.S. Government license to provide public access under the DOE Public Access Plan.

\section*{Data Availability}
The data that support the findings of this article are openly available~\cite{hernandez_morquecho_2026_19497772}.

\bibliography{biblio}% Produces the bibliography via BibTeX.

\newpage
\section{Supplementary Material}

\subsection{FLUKA Simulation}

\begin{figure*}[!ht]
   \centering
         \includegraphics[width=0.40\linewidth]{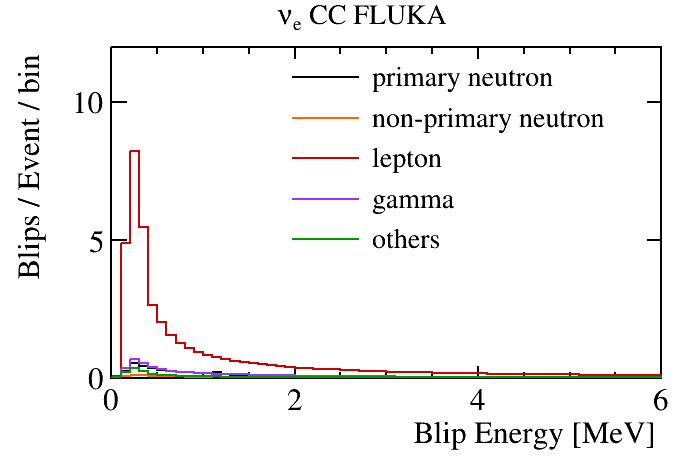}
         \includegraphics[width=0.40\linewidth]{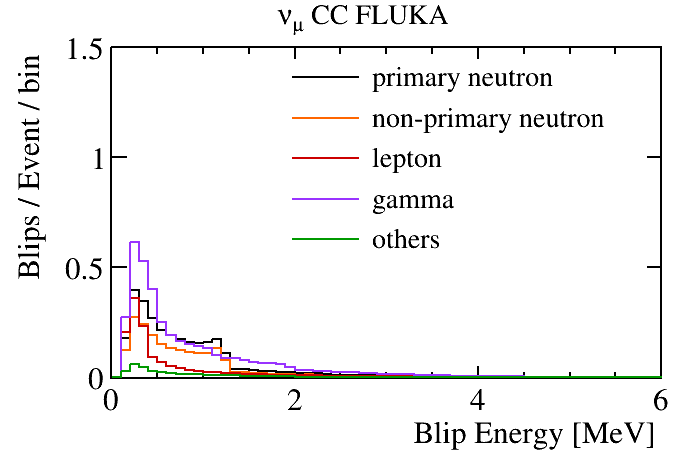}
         \includegraphics[width=0.40\linewidth]{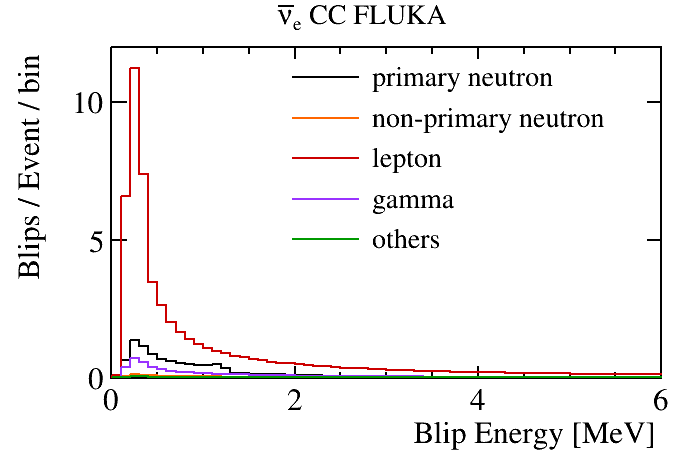}
         \includegraphics[width=0.40\linewidth]{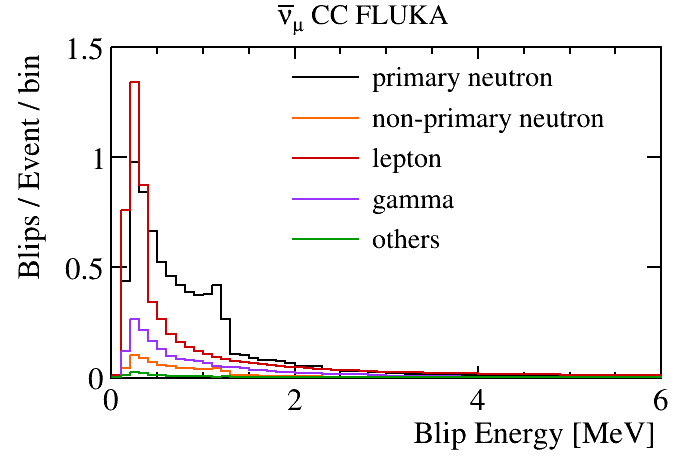}
         \caption{The distributions of the true blip energy for $\nu_e$ (top left), $\nu_\mu$ (top right), $\bar\nu_e$ (bottom left), and $\bar\nu_\mu$ (bottom right). No cut is applied.}
         \label{fig:blipE_true}
\end{figure*}

\begin{figure*}[!ht]
   \centering
         \includegraphics[width=0.40\linewidth]{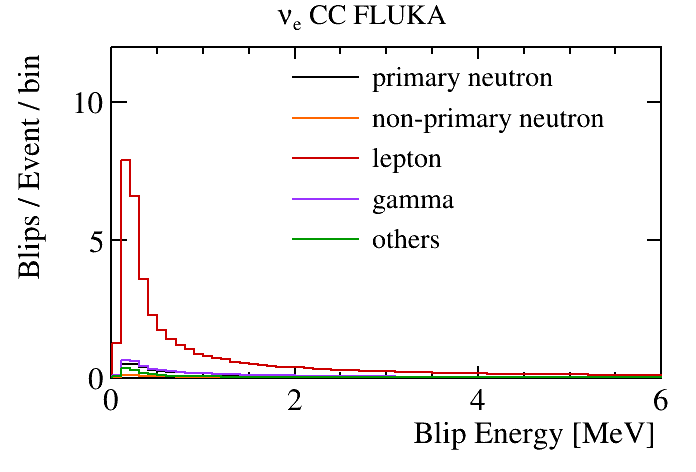}
         \includegraphics[width=0.40\linewidth]{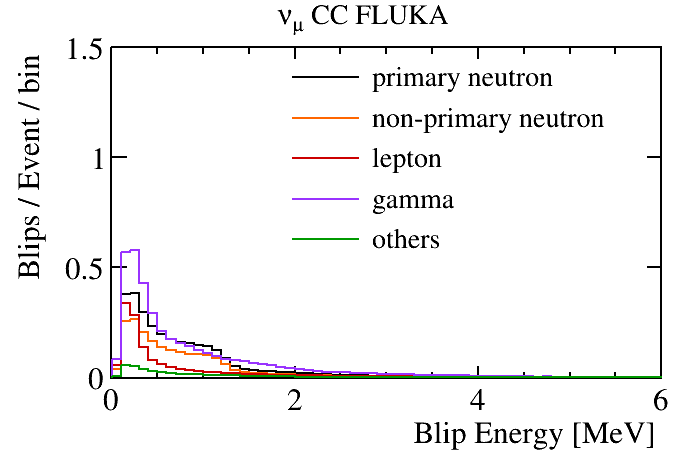}
         \includegraphics[width=0.40\linewidth]{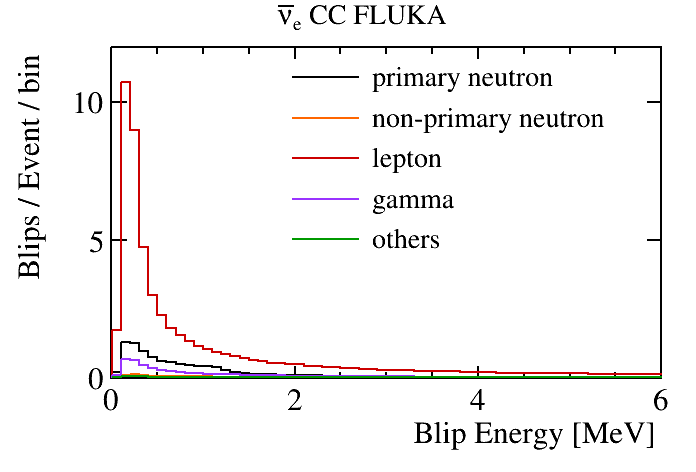}
         \includegraphics[width=0.40\linewidth]{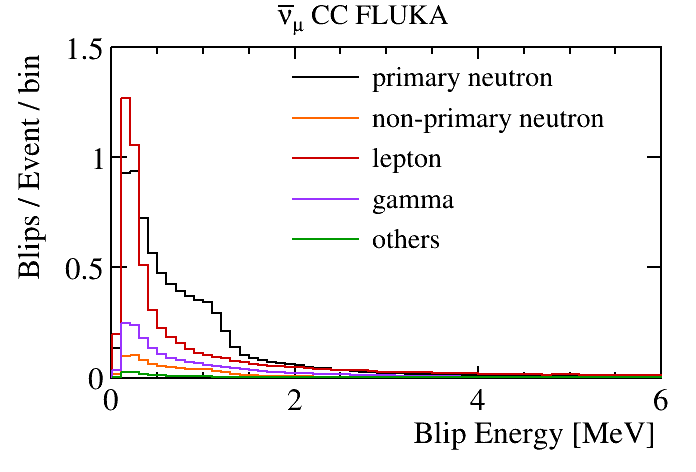}
         \caption{The distributions of the reconstructed blip energy for $\nu_e$ (top left), $\nu_\mu$ (top right), $\bar\nu_e$ (bottom left), and $\bar\nu_\mu$ (bottom right). No cut is applied.}
         \label{fig:blipE_nocut}
\end{figure*}

\begin{figure*}[!ht]
   \centering
         \includegraphics[width=0.40\linewidth]{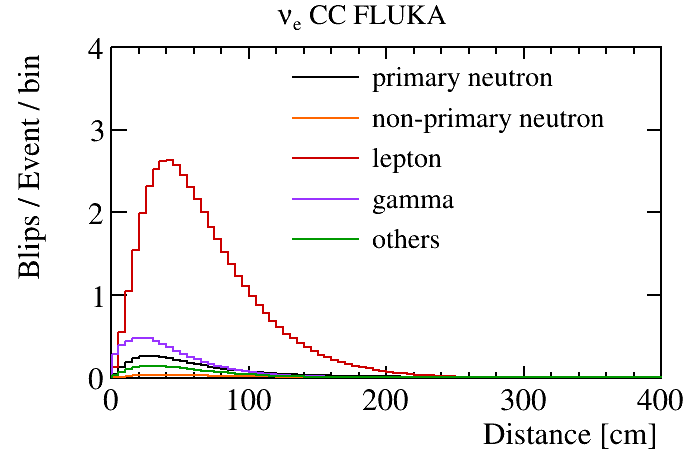}
         \includegraphics[width=0.40\linewidth]{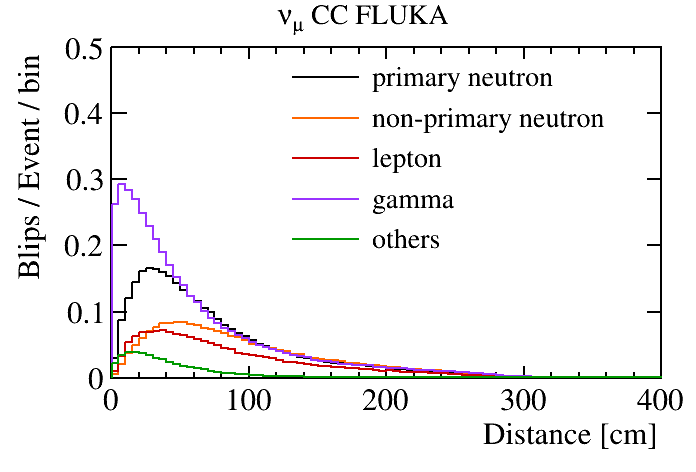}
         \includegraphics[width=0.40\linewidth]{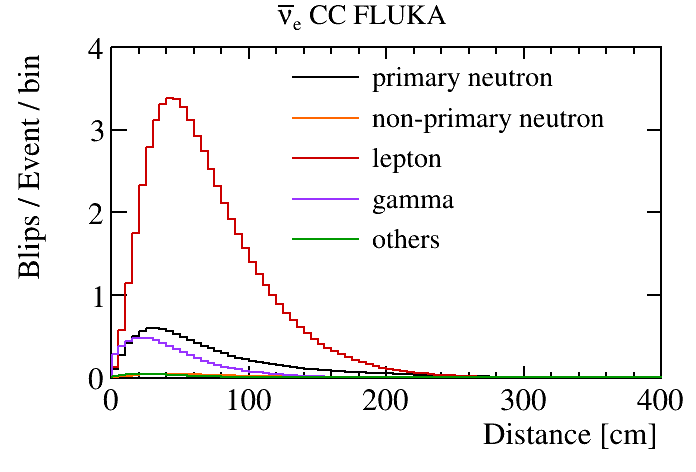}
         \includegraphics[width=0.40\linewidth]{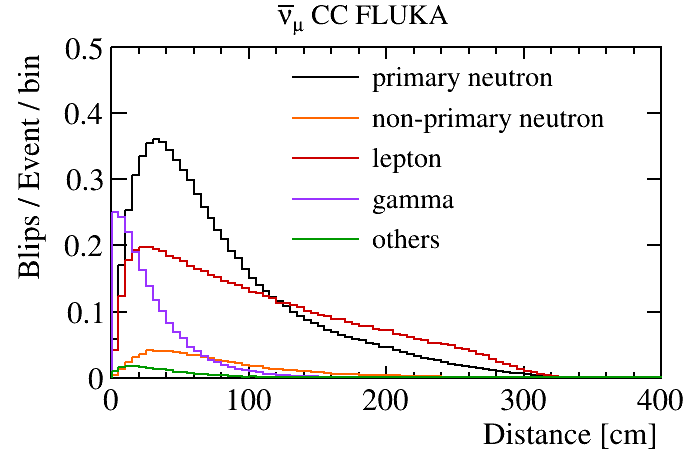}
         \caption{The distributions of the distance from the blip to the primary vertex for $\nu_e$ (top left), $\nu_\mu$ (top right), $\bar\nu_e$ (bottom left), and $\bar\nu_\mu$ (bottom right). No cut is applied.}
         \label{fig:blipD_nocut}
\end{figure*}

Before any selection, using the FLUKA simulation, the blip distribution in true energy is shown in Figure~\ref{fig:blipE_true}, and the reconstructed energy and distance to primary vertex is shown in Figure~\ref{fig:blipE_nocut} and Figure~\ref{fig:blipD_nocut}.

In summary, the blip selection for $\nu_\mu/\bar\nu_\mu$'s are:
\begin{itemize}
   \item PoCA $>$ \SI{20}{cm}
   \item The distance between blip vertex and primary neutrino interaction vertex is between \SIrange{20}{250}{cm}
   \item The distance between blip vertex and the end of the muon track is $>$ \SI{40}{cm}
   \item The reconstructed blip energy is $>$ \SI{0.6}{MeV}
\end{itemize}
, and the blip selection for $\nu_e/\bar\nu_e$'s are:
\begin{itemize}
   \item Cone angle $>$ \SI{45}{degree}
   \item The distance between blip vertex and primary neutrino interaction vertex is between \SIrange{20}{250}{cm}
   \item The reconstructed blip energy is $>$ \SI{0.6}{MeV}
\end{itemize}

\begin{figure*}[!ht]
   \centering
         \includegraphics[width=0.40\linewidth]{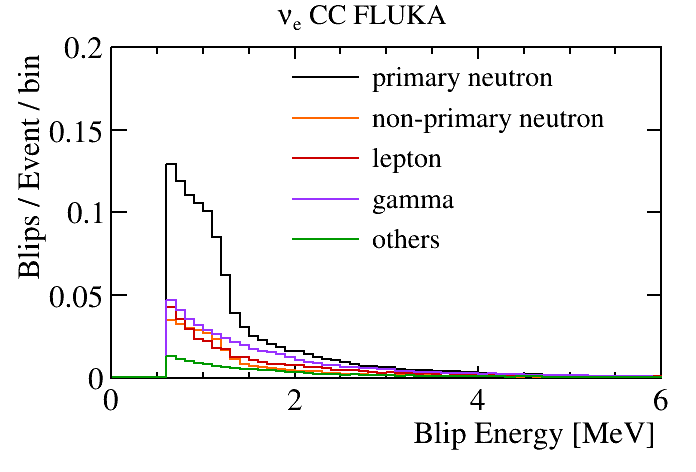}
         \includegraphics[width=0.40\linewidth]{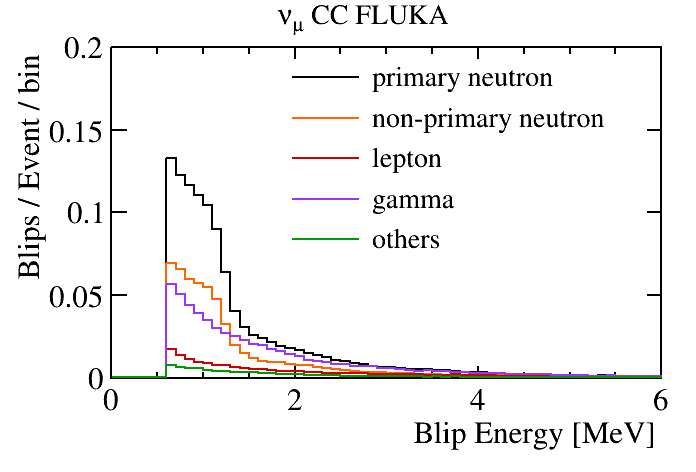}
         \includegraphics[width=0.40\linewidth]{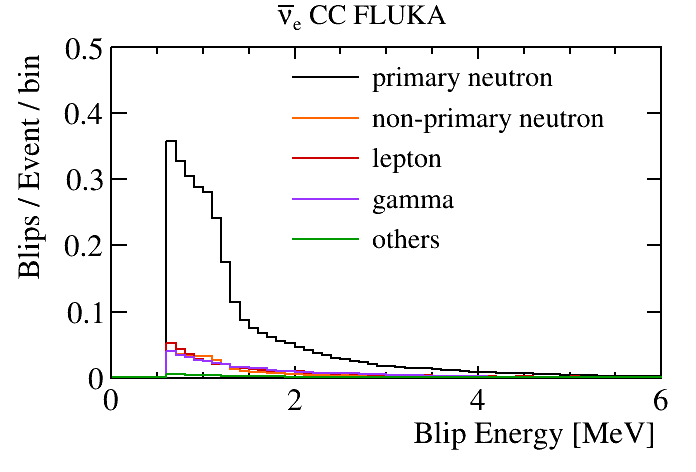}
         \includegraphics[width=0.40\linewidth]{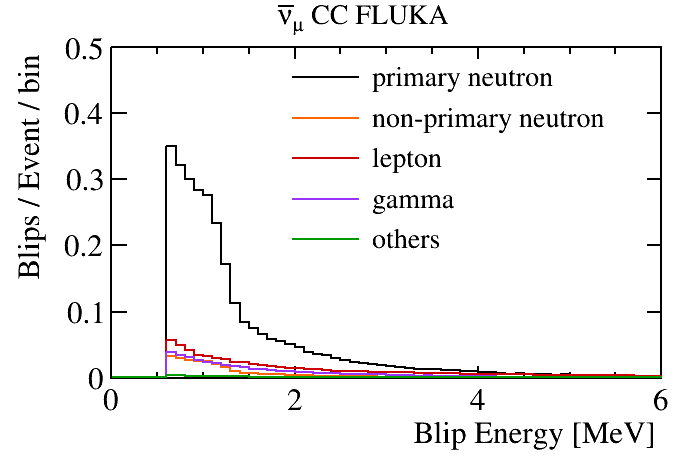}
         \caption{The distributions of the reconstructed blip energy for $\nu_e$ (top left), $\nu_\mu$ (top right), $\bar\nu_e$ (bottom left), and $\bar\nu_\mu$ (bottom right). All cuts are applied.}
         \label{fig:blipE_allcut}
\end{figure*}

\begin{figure*}[!ht]
   \centering
         \includegraphics[width=0.40\linewidth]{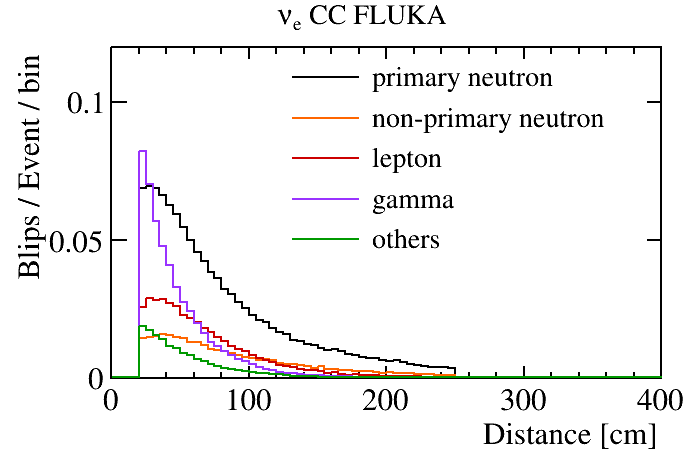}
         \includegraphics[width=0.40\linewidth]{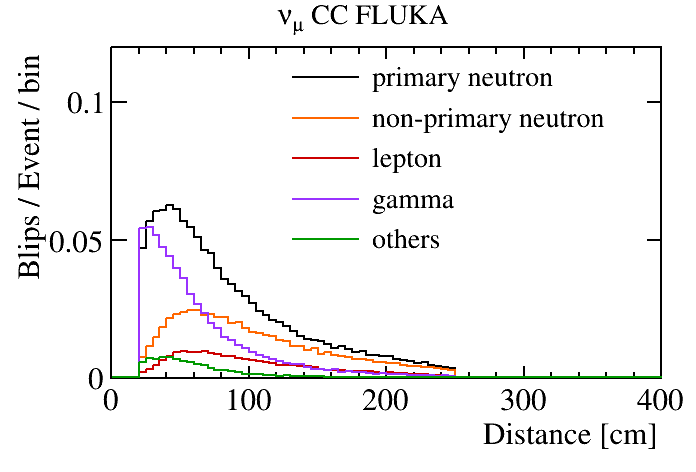}
         \includegraphics[width=0.40\linewidth]{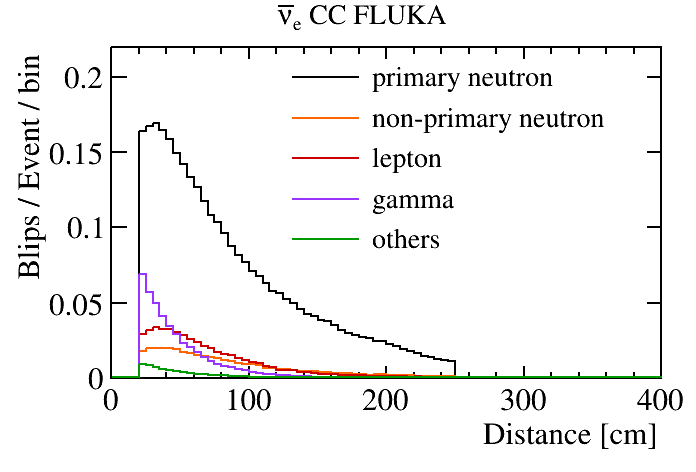}
         \includegraphics[width=0.40\linewidth]{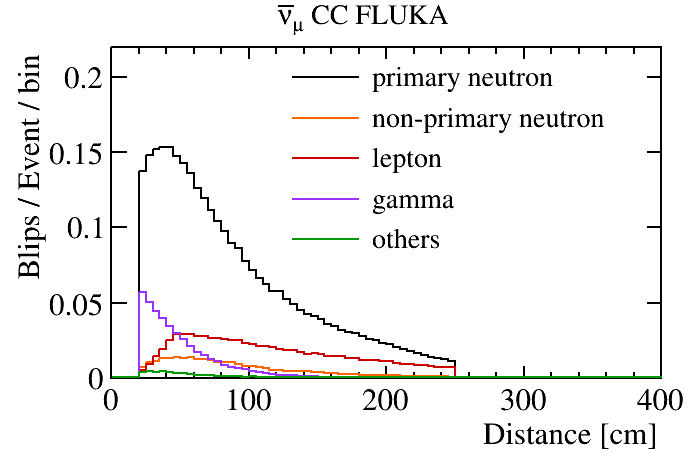}
         \caption{The distributions of the distance from the blip to the primary vertex for $\nu_e$ (top left), $\nu_\mu$ (top right), $\bar\nu_e$ (bottom left), and $\bar\nu_\mu$ (bottom right). All cuts are applied.}
         \label{fig:blipD_allcut}
\end{figure*}

After all the selections, the remaining blip distribution in reconstructed energy and distance to primary vertex is shown in Figure~\ref{fig:blipE_allcut} and Figure~\ref{fig:blipD_allcut}.

\subsection{Geant4 Simulation}

\begin{figure*}[!ht]
   \centering
         \includegraphics[width=0.40\linewidth]{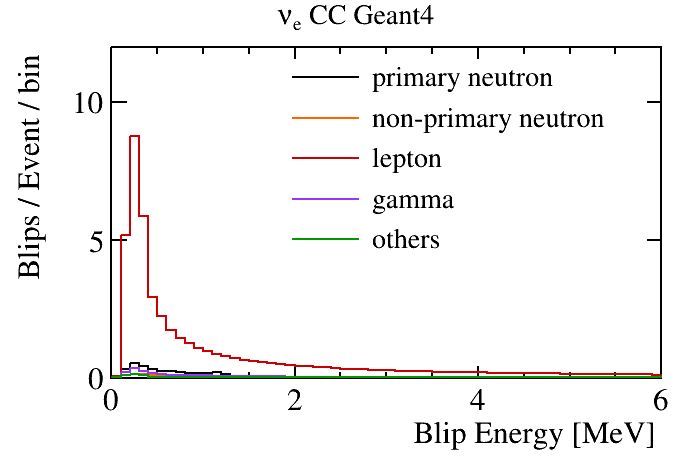}
         \includegraphics[width=0.40\linewidth]{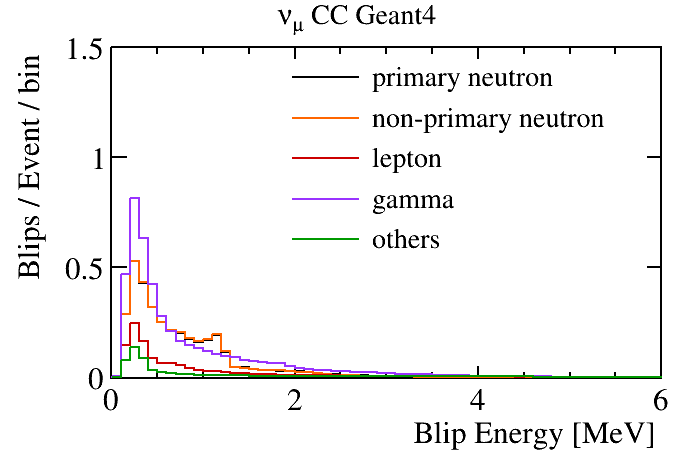}
         \includegraphics[width=0.40\linewidth]{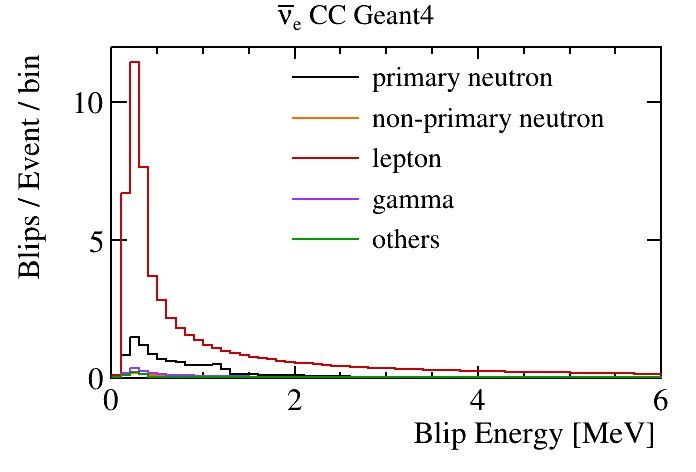}
         \includegraphics[width=0.40\linewidth]{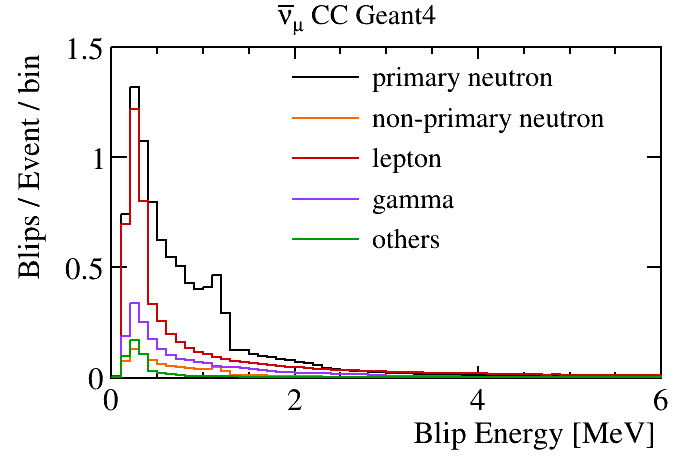}
         \caption{The distributions of the true blip energy for $\nu_e$ (top left), $\nu_\mu$ (top right), $\bar\nu_e$ (bottom left), and $\bar\nu_\mu$ (bottom right). No cut is applied.}
         \label{fig:G4blipE_true}
\end{figure*}

\begin{figure*}[!ht]
   \centering
         \includegraphics[width=0.40\linewidth]{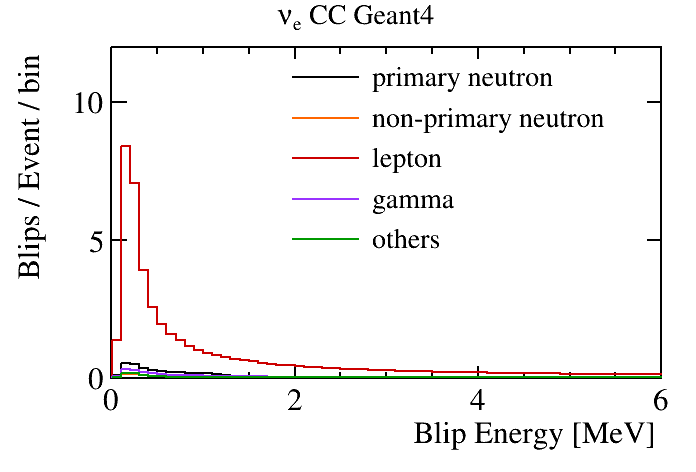}
         \includegraphics[width=0.40\linewidth]{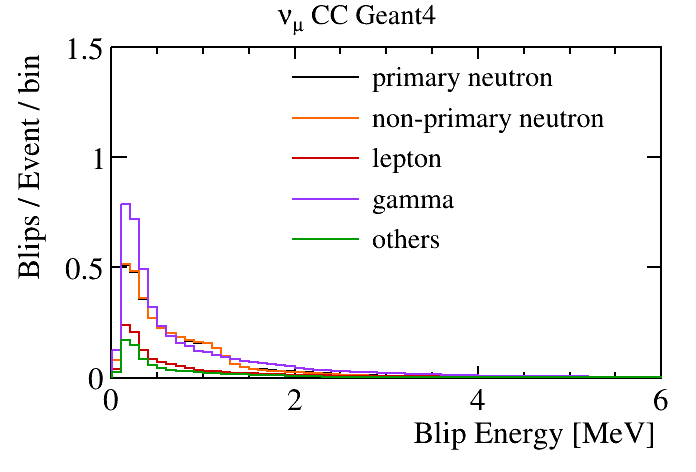}
         \includegraphics[width=0.40\linewidth]{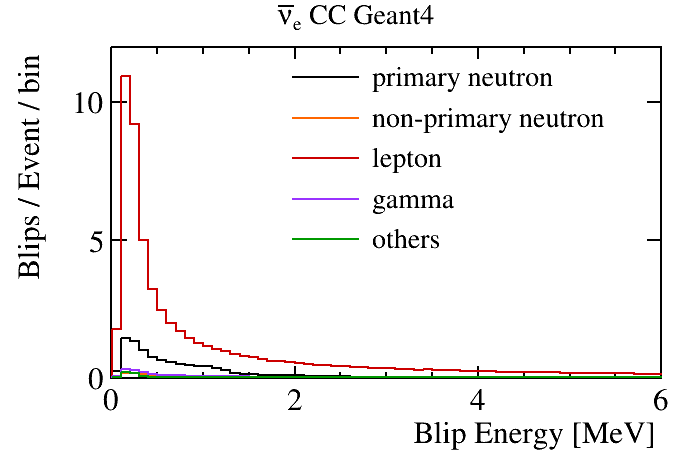}
         \includegraphics[width=0.40\linewidth]{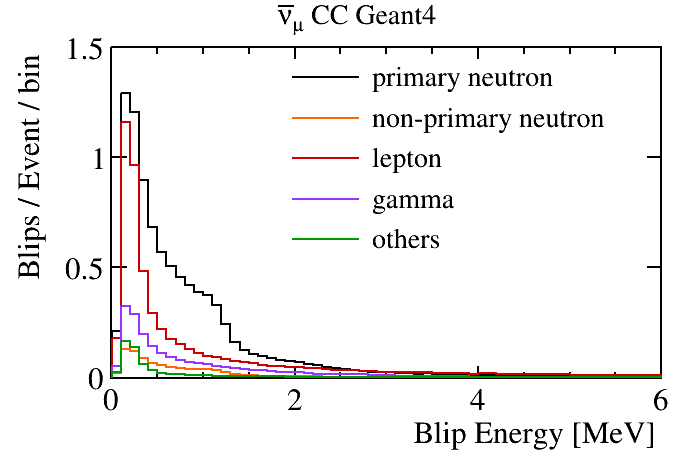}
         \caption{The distributions of the reconstructed blip energy for $\nu_e$ (top left), $\nu_\mu$ (top right), $\bar\nu_e$ (bottom left), and $\bar\nu_\mu$ (bottom right). No cut is applied.}
         \label{fig:G4blipE_nocut}
\end{figure*}

\begin{figure*}[!ht]
   \centering
         \includegraphics[width=0.40\linewidth]{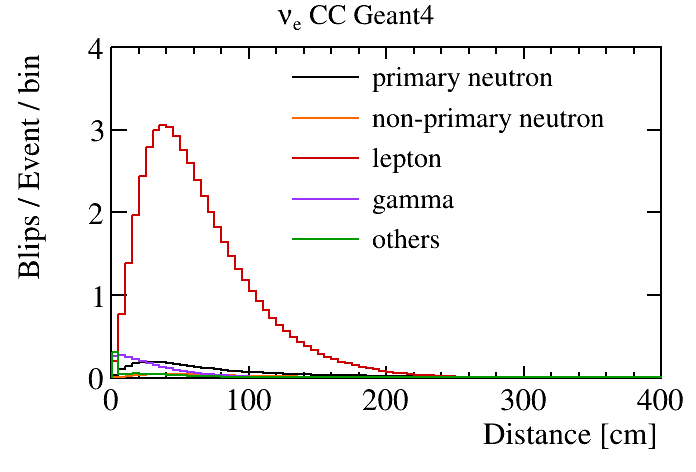}
         \includegraphics[width=0.40\linewidth]{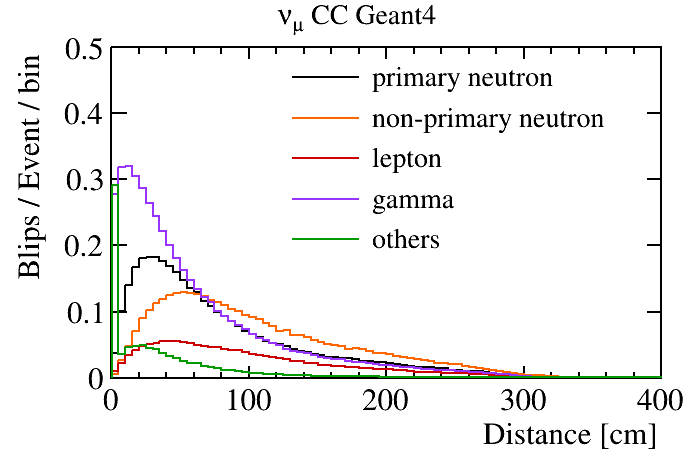}
         \includegraphics[width=0.40\linewidth]{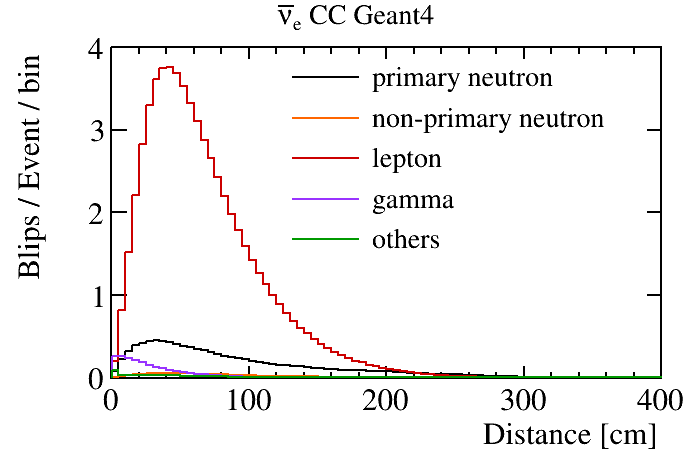}
         \includegraphics[width=0.40\linewidth]{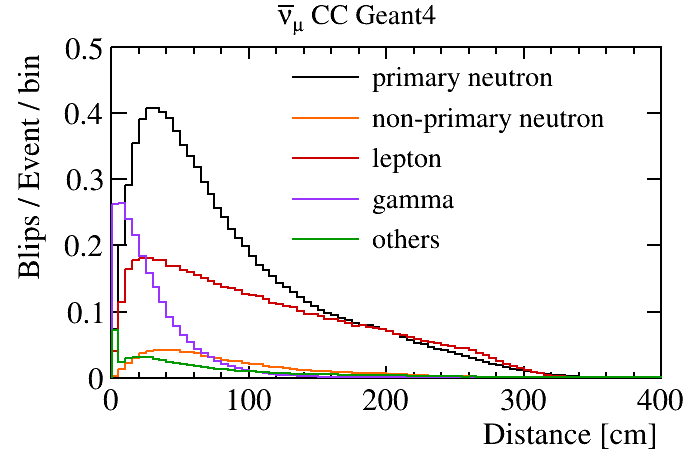}
         \caption{The distributions of the distance from the blip to the primary vertex for $\nu_e$ (top left), $\nu_\mu$ (top right), $\bar\nu_e$ (bottom left), and $\bar\nu_\mu$ (bottom right). No cut is applied.}
         \label{fig:G4blipD_nocut}
\end{figure*}

Using the alternative Geant4 simulation, the blip distribution in true energy is shown in Figure~\ref{fig:G4blipE_true}, and the reconstructed energy and distance to primary vertex before any selection is shown in Figure~\ref{fig:G4blipE_nocut} and Figure~\ref{fig:G4blipD_nocut}.

\begin{figure*}[!ht]
   \centering
         \includegraphics[width=0.40\linewidth]{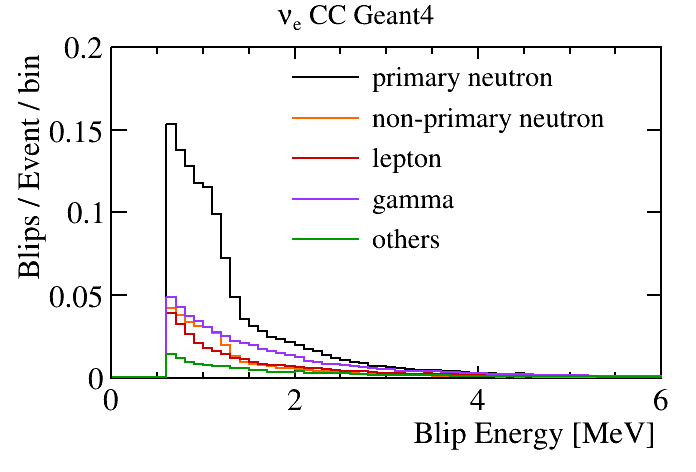}
         \includegraphics[width=0.40\linewidth]{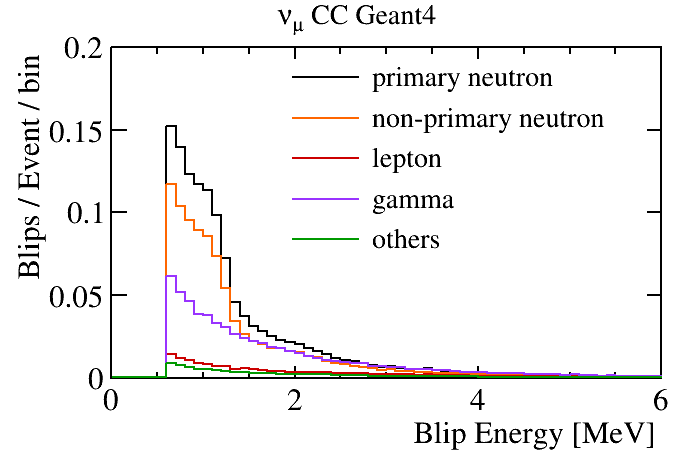}
         \includegraphics[width=0.40\linewidth]{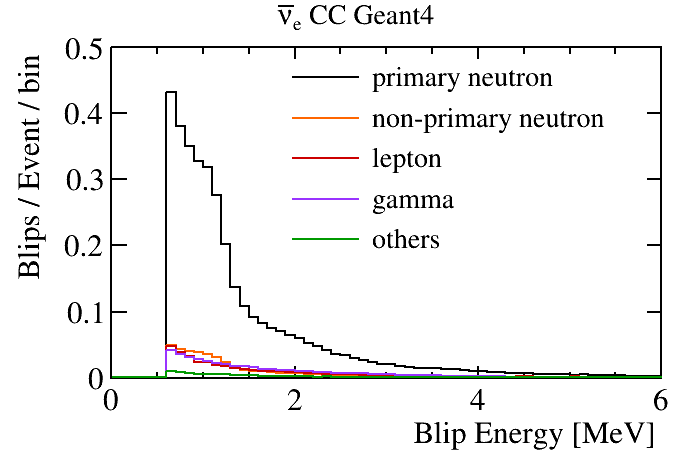}
         \includegraphics[width=0.40\linewidth]{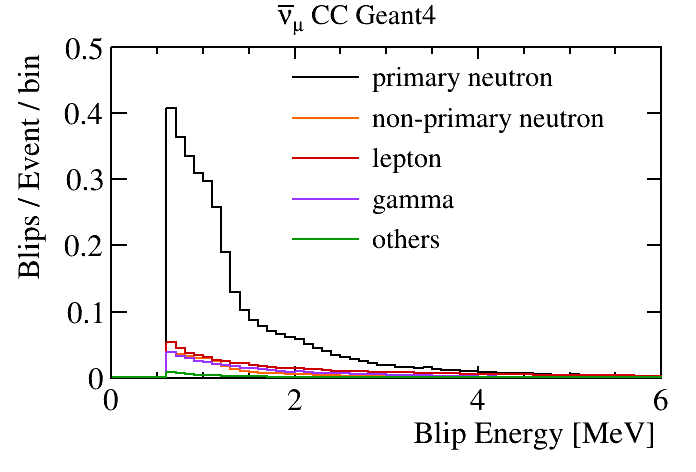}
         \caption{The distributions of the reconstructed blip energy for $\nu_e$ (top left), $\nu_\mu$ (top right), $\bar\nu_e$ (bottom left), and $\bar\nu_\mu$ (bottom right). All cuts are applied.}
         \label{fig:G4blipE_allcut}
\end{figure*}

\begin{figure*}[!ht]
   \centering
         \includegraphics[width=0.40\linewidth]{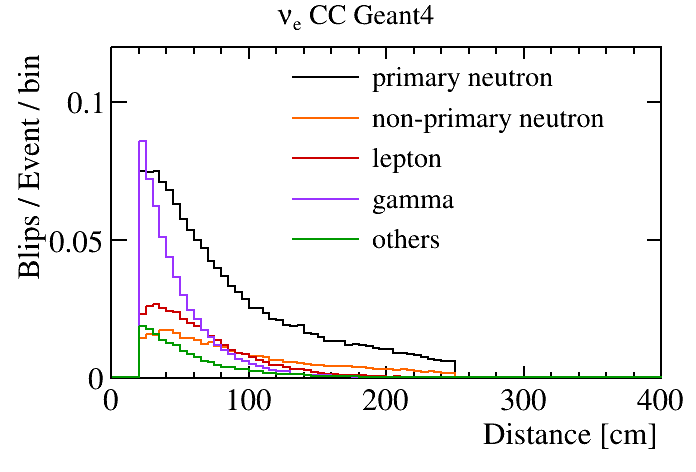}
         \includegraphics[width=0.40\linewidth]{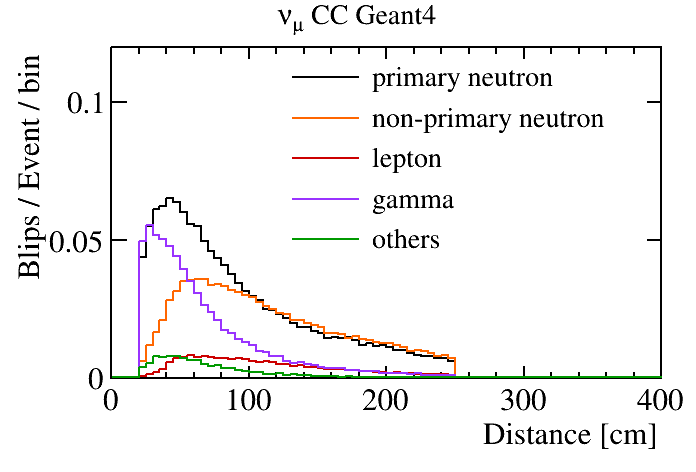}
         \includegraphics[width=0.40\linewidth]{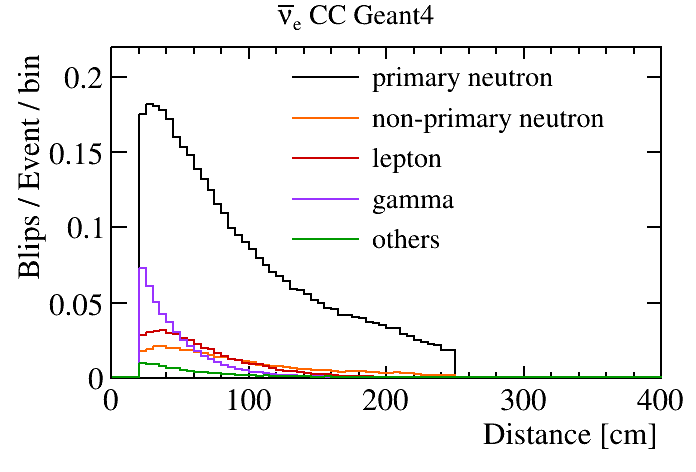}
         \includegraphics[width=0.40\linewidth]{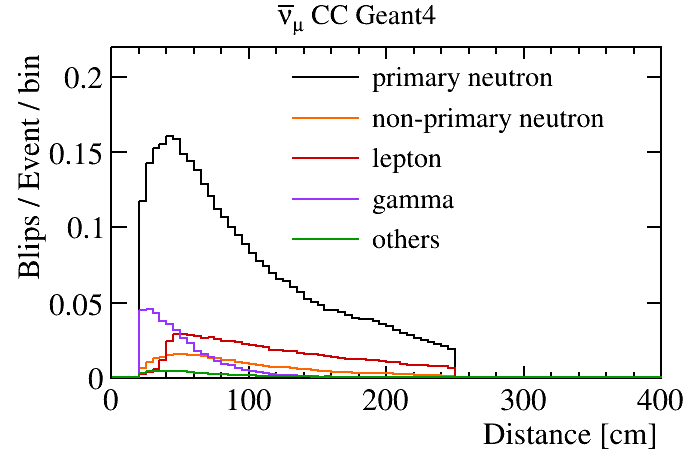}
         \caption{The distributions of the distance from the blip to the primary vertex for $\nu_e$ (top left), $\nu_\mu$ (top right), $\bar\nu_e$ (bottom left), and $\bar\nu_\mu$ (bottom right). All cuts are applied.}
         \label{fig:G4blipD_allcut}
\end{figure*}

After all the selections, the remaining blip distribution in reconstructed energy and distance to primary vertex is shown in Figure~\ref{fig:G4blipE_allcut} and Figure~\ref{fig:G4blipD_allcut}.

\end{document}